\documentclass[preprint]{revtex4}

\usepackage{graphicx}
\usepackage{booktabs}
\usepackage{amssymb}
\usepackage{amsmath}
\usepackage{color}

\begin{document}
\include{commands}

\title{Periodic Poisson Solver for Particle Tracking}

\author{M.~Dohlus, Ch.~Henning\\
       \it Deutsches Elektronen-Synchrotron (DESY), D-22607 Hamburg, Germany}

\begin{abstract}
A method is described to solve the Poisson problem for a three dimensional source distribution that is periodic into one direction. Perpendicular to the direction of periodicity a free space (or open) boundary is realized. In beam physics, this approach allows to calculate the space charge field of a continualized charged particle distribution with periodic pattern.

The method is based on a particle mesh approach with equidistant grid and fast convolution with a Green's function. The periodic approach uses only one period of the source distribution, but a periodic extension of the Green's function.

The approach is numerically efficient and allows the investigation of periodic- and pseudo-periodic structures with period lengths that are small compared to the source dimensions, for instance of laser modulated beams or of the evolution of micro bunch structures. Applications for laser modulated beams are given.
\end{abstract}

\maketitle

\section{Introduction}
The fundamental problem of the analysis of beam dynamics is to solve the equation
of motion for a multi particle system ($N \sim 10^3 \cdots 10^9$) in presence of
external forces (caused by magnets, cavities etc.) and self forces. A particular
difficulty is to determine the self-fields of the dynamic many-particle system. The
coarsest approximation to solve this problem is the space charge or Poisson approach. 
It assumes that all particles are in uniform motion with identical velocity vector.
Therefore the problem is equivalent to an electrostatic problem, and in principle one
has to add for each observer particle the one over distance contribution from all other
sources. This point-to-point interaction would allow the slightly better approach
of individual uniform motion per particle, but its effort scales as $N^2$. This
is avoidable by a continualization so that the point particles are represented by
a smooth density function. Additionally the continualization solves a second
problem that is usually not in the focus of interest: the near scattering of particles
that approach much closer than the averaged particle distance. A proper tracking with
near scattering needs not only the point-to-point interaction but also a very
fine time step related to the minimal distance between particles.

Continuous density functions and fields are numerically represented either by
superpositions of basis functions (finite element methods) or by mesh methods.
In this report we use a three dimensional equidistant mesh for binning and compute
the static potential by a fast convolution with a Green's function as described in
\cite{Hockney, Qiang}.
This Green's function fulfills free space (or open) boundary conditions.
Also the electric field is calculated on the mesh, by numeric
differentiation. The field is interpolated to the particle positions and transformed
into a collective force as it appears in the equation of motion. The numeric integration
of the equation of motion is not described here.

The Poisson-particle-mesh approach is used widely and successfully in beam dynamics
for instance in straight sections with constant energy, but also in particle sources or accelerating sections
 and even in sections with weak curvature. Sometimes the particle distribution has a
fine sub-structure which has to be taken into account, for instance the particles
perform plasma oscillations
\cite{Lawson,GSSY_plasma,SPIE_Prag},
a micro modulation builds up (micro-bunch-instability) or they interact in
an undulator with a laser. Such high resolution computations need very fine meshes
and high particle (or macro-particle) numbers so that they become numerically very expensive.

This numerical effort can be dramatically reduced when the approach assumes and utilizes periodicity: only the particles of one period have to be tracked and meshed. We describe such an approach where the periodicity of the source distribution is interchanged with the periodicity of the Green's function. Except for the calculation of the Green's function, the established method (Poisson-particle-mesh with equidistant mesh and fast convolution) can be used completely unchanged. We consider periodicity into one direction. The direction is arbitrary and needs not to coincide with a mesh axis. Perpendicular to this direction free space (or open) boundary conditions are realized.

The report is organized as follows. Section II is about general aspects of periodicity.
In section III we describe the approach, first the standard method (Poisson, particle-mesh,
convolution with Green's function), then its modification with periodic Green's function.
In section IV we demonstrate the method with examples: the FLASH seeding section
\cite{FLASH, Hacker, SASE_Sup}
with different particle densities and modulations, and parasitic heating after the LCLS laser heater
\cite{LCLS, Huang LH}.
For comparison of field calculation methods, we did tracking simulations with Astra
\cite{Floettmann}
and the non-periodic and periodic implementation of the method described in this report, called QField.


\section{Periodicity}
Some aspects of spatial periodicity and pseudo-periodicity are illustrated in Fig.~\ref{per1}. The particles in panel (a) are randomly distributed while panel (b) shows a periodically repeated set $S_{\rm p}$ of random particles. Distribution (a) is caused by applying a periodic transport map to random particles that are uniformly distributed into the direction of periodicity. Strictly periodic distributions as in (b) are used to approximate pseudo-periodic distributions as in (a).

For the strictly periodic approach we consider only a set $S_{\rm p}$ of non-repeated particles. As it can be seen in panels (b-e) there are many sets $S_{\rm p}$ that describe the same periodic distribution. It is easy to manipulate such sets by shifting particles by a multiple of the repetition vector ${\bf r}_{\rm p}$. The sets in (c) and (d) are chosen so that all particles are in a volume slice, where the front plane corresponds to the back plane shifted by ${\bf r}_{\rm p}$. These slices might have any orientation (or shift), but there are orientations so that the volume of a box (or sphere) that includes all particles is minimal. The distribution in (b) are particles from one volume slice, that have been mapped by a periodic transport map. (Particles with nearly the same spatial coordinates, but with different momenta may be mapped to very different spatial coordinates.) For numerical reasons it might be advantageous to use sets without sudden change of density, as it can be seen in panel (e). This set has been composed similar as (c), but some particles close to the left boundary are shifted by ${\bf r}_{\rm p}$ (beyond the right boundary) and some particles close to the right boundary are shifted into the opposite direction. The probability of shifting depends on the distance to the boundary.

To keep periodicity, the set $S_{\rm p}$ has to represent a distribution that is periodic in six dimensional phase space $X=\left[x,y,z,p_x,p_y,p_z\right]$. Periodically shifted particles are at $X+nP$, with $P$ the vector of six dimensional periodicity. The required property for the transport $T(X) \rightarrow Y$ is $T(X+nP) \rightarrow T(X)+nT(P)$. This is for instance fulfilled if only one spatial component of $P$ is non-zero and $T$ is a linear function of that component. 

Real distributions are always pseudo-periodic since they are finite. Sometimes only a small part of a (long) bunch behaves periodically, as for instance in Figs.~\ref{Cur_250keV_before_after_CH3}, \ref{EvZ_250keV_189p4_204p2_Astra_xyz2}. The phase space volume covered by the beam is usually small and the transport can be linearized, or at least locally (for one ore few periods) linearized. Therefore the Poisson approach can be used for distributions with finite extension in momenta space, although the uniform motion approach assumes identical momentum of all particles. In the same sense the periodical Poisson approach is applicable to distributions with non-zero momentum components of $P$, although this would mean particles with arbitrarily large momentum. An example is a bunch with energy chirp and micro modulation.

The Coulomb force drops with $1/r$ and the effect of a dipole perturbation drops with $1/r^2$. Therefore space charge effects are very localized and non-resonant. Pseudo-periodic behavior can be observed even for distributions with only few periods, as it can be seen in the first example of section IV.

\section{Method}
\subsection{Poisson Approach and Lorentz Force}
The electro-magnetic field caused by a set $S$ of discrete point particles with charge $q_i$, position ${\bf r}_i(t)$ and velocity ${\bf v}_i(t)$ is estimated based on the assumption of collective uniform motion (of all particles) with the velocity  ${\bf v}_{\rm c}$.
The particle positions are represented by orthonormal coordinates ${\bf r}_i=x_i {\bf e}_{\perp 1}+y_i {\bf e}_{\perp 2}+z_i {\bf e}_{\rm c}$, wherein the $z$-axis points in the direction of motion ${\bf e}_{\rm c}={\bf v}_{\rm c}/\| {\bf e}_{\rm c} \|$ and the directions ${\bf e}_{\perp 1}$ and ${\bf e}_{\perp 2}$ are perpendicular.
After Lorentz transformation to the rest frame, the coordinates are ${\bf r}_{i,{\rm c}}=x_i {\bf e}_{\perp 1}+y_i {\bf e}_{\perp 2}+z_i \gamma_{\rm c}{\bf e}_{\rm c}$ with $\gamma_{\rm c}=1/\sqrt{1-\beta_{\rm c}^2}$ and $\beta_{\rm c}=v_{\rm c}/c=\| {\bf v}_{\rm c} \|/c$.
The electrostatic field may be computed from point-to-point interaction as
\begin{equation}
{\bf E}_{i,{\rm c}}=\frac{1}{4 \pi \varepsilon_0} \sum_{j \in S_i} q_j
\frac{{\bf r}_{i,{\rm c}}-{\bf r}_{j,{\rm c}}}{\| {\bf r}_{i,{\rm c}}-{\bf r}_{j,{\rm c}} \|^3}
\: \: ,
\end{equation}
with excluded self interaction. Therefore $S_i$ is the subset of all particles without particle $i$. The
continualization replaces the particle distribution $\{q_i,{\bf r}_{i,{\rm c}}\}$ by a continuous charge density
function $\rho_c({\bf r})$ and the electrostatic field
\begin{equation}
{\bf E}_c({\bf r}) = - \nabla \phi_c({\bf r})
\end{equation}
is the negative gradient of the electrostatic potential
\begin{equation}
\phi_c({\bf r})=\frac{1}{4 \pi \varepsilon_0} \int dV' \times \frac{\rho_c({\bf r}')}{\| {\bf r}-{\bf r}'\|}
\label{Poisson_integral}
\: \: .
\end{equation}
Finally we transform back to the moving frame
\begin{eqnarray}
{\bf E}_{i}&=&\gamma_{\rm c} \left({\bf e}_{\perp 1} \left({\bf e}_{\perp 1} \cdot {\bf E}_{i,{\rm c}}\right)+{\bf e}_{\perp 2} \left({\bf e}_{\perp 2} \cdot {\bf E}_{i,{\rm c}}\right)\right)+{\bf e}_{\rm c} ({\bf e}_{\rm c} \cdot {\bf E}_{i,{\rm c}})\\
{\bf B}_{i}&=&c^{-2} {\bf v}_{\rm c} \times {\bf E}_{i}
\end{eqnarray}
and calculate the Lorentz force
\begin{equation}
{\bf F}_{i}=e \left( {\bf E}_{i} +{\bf v}_{i}\times {\bf B}_i\right)
=e \left( {\bf E}_{i} +c^{-2}{\bf v}_{i}\times {\bf v}_{\rm c} \times {\bf E}_{i} \right)
\: \: .
\end{equation}

\subsection{Particle-Mesh-Method}
A simple and robust continualization method is to approximate the discrete charges by a continuous distribution that is piecewise constant in the cells of an equidistant mesh. The mesh points ${\bf r}_{\rm m}(j,k,l)=j h_x {\bf e}_{\perp 1}+k h_y {\bf e}_{\perp 2}+l h_z {\bf e}_{\rm c}$ are in the middle of the cells and the charges per cell are $q_{\rm c}(j,k,l)$. For simplicity we have chosen the direction of motion ${\bf e}_{\rm c}$ to coincide with one mesh axis. This has numerical advantages, as the resolution requirement in longitudinal direction is usually very different from that into transverse directions, but it is not necessary. The choice of the perpendicular directions ${\bf e}_{\perp 1}$, ${\bf e}_{\perp 2}$ is free.

The electrostatic potential caused by the charge density $\rho=q_c(0,0,0)/(h_x h_y h_z)$ in the cell which includes the origin, is 
\begin{equation}
\phi({\bf r})=\frac{\rho}{4 \pi \varepsilon_0} G({\bf r}) 
\end{equation}
with the Green's function
\begin{equation}
G(x {\bf e}_{\perp 1}+y {\bf e}_{\perp 2}+z {\bf e}_{\rm c})=\int_{-h_x/2}^{h_x/2}\int_{-h_y/2}^{h_y/2}\int_{-h_z/2}^{h_z/2} \frac{dx'dy'dz'}{\sqrt{(x-x')^2+(y-y')^2+(z-z')^2}}
\label{greens_function_integral}
\: \: .
\end{equation}
This integral can be solved as \cite{Qiang}:
\begin{equation}
G(x {\bf e}_{\perp 1}+y {\bf e}_{\perp 2}+z {\bf e}_{\rm c}) = \sum_{i,j,k\in \{-1,1\}} ijk \: H(x+ih_x/2,y+jh_y/2,z+kh_z/2) \\
\label{greens_function}
\: \: ,
\end{equation}
with the anti-derivative
\begin{eqnarray}
H(x,y,z)&=&-\frac{x^2}{2}\arctan\frac{yz}{xr}-\frac{y^2}{2}\arctan\frac{zx}{yr}-\frac{z^2}{2}\arctan\frac{xy}{zr}
\nonumber \\
&&+xy\ln(z+r)+yz\ln(x+r)+zx\ln(y+r)
\label{anti_derivative_of_greens_function}
\:\:,
\end{eqnarray}
and $r=\sqrt{x^2+y^2+z^2}$. The asymptotic behavior
\begin{equation}
G({\bf r}) \rightarrow G_a({\bf r}) = h_x h_y h_z / r 
\label{approximated_greens_function}
\end{equation}
is obvious from Eq. (\ref{greens_function_integral}).

To compute the potential at all mesh points $\phi_{\rm m}(j,k,l)$, caused by all mesh charges, one has to
perform the summation
\begin{equation}
\phi_{\rm m}(j,k,l)=\frac{1}{4 \pi \varepsilon_0} \frac{1}{h_x h_y h_z} \sum_{j',k',l'}  q_{\rm c}(j',k',l') g(j-j',k-k',l-l')
\: \: ,
\end{equation}
with $g(j,k,l)$ the Green's function at the mesh points ${\bf r}_{\rm m}(j,k,l)$.
This can be done efficiently by a three-dimensional fast convolution. The components of the electric
field are determined as difference quotients (f.i. $E_c = [ \phi_m(j,k,l)-\phi_m(j,k,l+1) ]/h_z$). These components, that are allocated on shifted meshes (as ${\bf r}_{\rm m}(j,k,l+1/2)$), are interpolated to the
Lorentz transformed particle positions ${\bf r}_{i,{\rm c}}$ to find the fields ${\bf E}_{i,{\rm c}}$ in the rest frame.

For an efficient computation of the Green's function on the mesh, one utilizes the asymptotic behavior for points far from the origin and calculates only near points by Eq. (\ref{greens_function}). Therefore one calculates first the anti derivative $H([j-1/2] h_x,[k-1/2] h_y,[l-1/2] h_z)$ in one octant ($j,k,l \ge 0$) of the near-volume and then the Green's function itself. The field in other octants follows from mirror symmetry. Technically, the change to the asymptotic approximation is realized with a smooth switch function
\begin{equation}
S(x)=\left\{ \begin{array}{ll} 0 & \mbox{if $x<C_1$}  \\
                               1 & \mbox{if $x>C_2$} \\
                               0.5-0.5 \cos\left(\pi\frac{x-C_1}{C_2-C_1}\right) & \mbox{otherwise}                    
             \end{array}
     \right.
\end{equation}
as ``switched'' Green's function
\begin{equation}
G_s({\bf r})=G({\bf r})+\left( G_a({\bf r}) -G({\bf r}) \right) S(r/\max(h_x,h_y,h_z))
\label{switched_greens_function}
\: \: .
\end{equation} 
The volume of pure asymptotic behaviour follows from the condition
\begin{equation}
    r > C_2 \max(h_x,h_y,h_z)
\: \: .
\label{criterion_for_far}
\end{equation}

\begin{figure}
\includegraphics[width=0.90\textwidth]{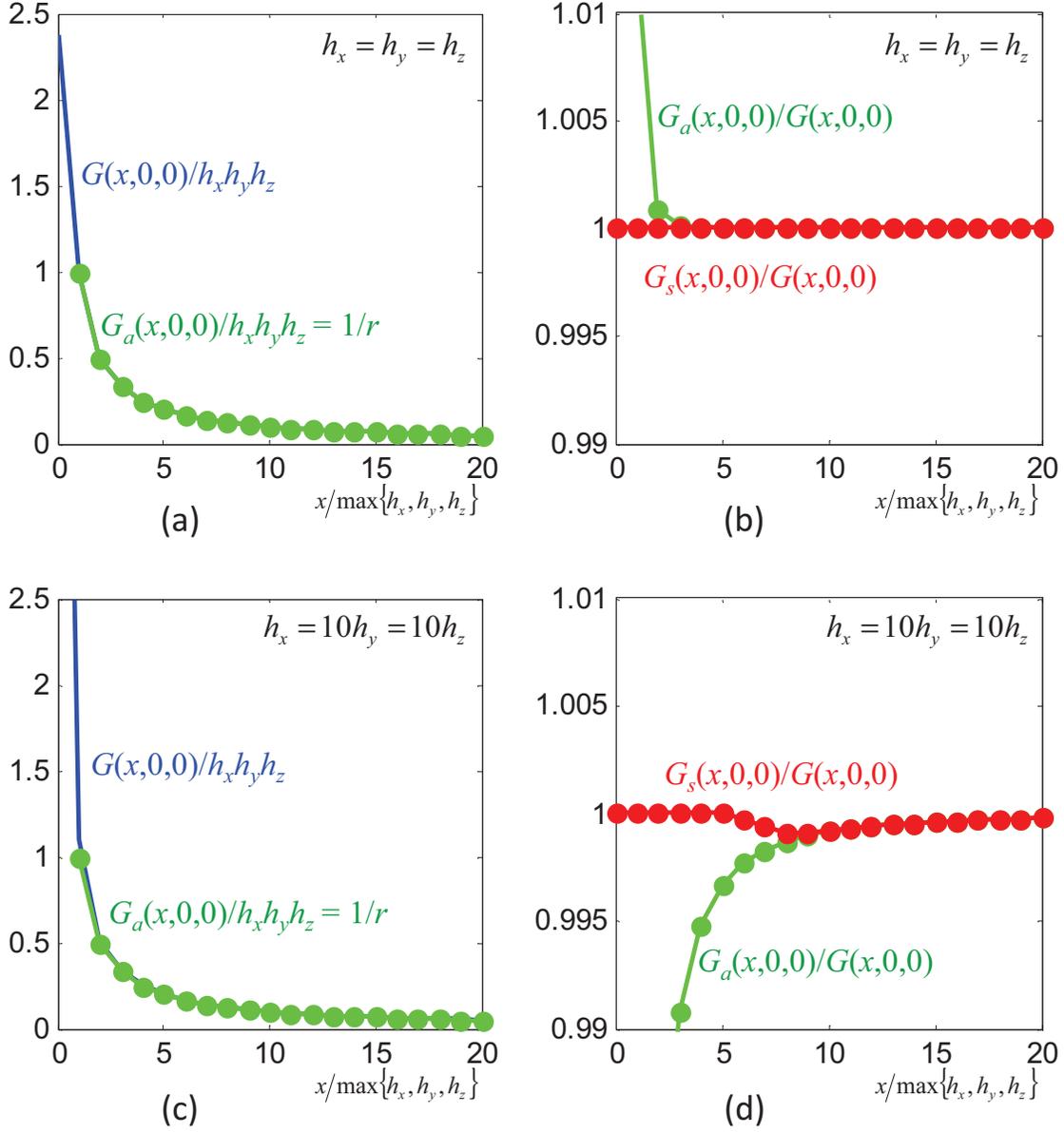} 
\caption{The Green's function $G$, the asymptotic approximation $G_a$ and the ``switched'' approximation $G_s$ for switch parameters $C_2=2C_1=10$: (a) and (c) comparison of $G$ and
its asymptotic approximation $G_p$; (b) and (d) ratios $G_a/G$ and $G_s/G$; (a) and (b) for
cubic mesh cells; (c) and (d) for mesh cells with large aspect ratio ($h_x=10h_y=10h_z$). The Green's
funtion is only calculated  for discrete mesh points ($x=n h_x$).}
\label{g_approx}
\end{figure}

In Fig.~\ref{g_approx} the Green's function is compared with the asymptotic approximation, and the ratios $G_a/G$ and  $G_s/G$ are plotted for cubic mesh cells ($h_x=h_y=h_z$) and for mesh cells with large aspect ratio ($h_x=10h_y=10h_z$). The relative deviation of the asymptotic function is below 0.001 for $x/max(h_x,h_y,h_z)>10$. Note that the case of infinite aspect ratio ($max(h_y,h_z)/h_x \rightarrow 0$) is very similar to the case $h_x=10h_y=10h_z$ in the volume $r \ge 2 h_x$. The switch function is calculated for the parameters $C_2=2C_1=10$. The relative deviation of the ``switched'' function is below 0.001 for the hole volume.

\subsection{Periodic Source and Periodic Green's Function}
Each particle of the set $S_{\rm p}$ is infinitely repeated with the spatially periodic shift ${\bf r}_{\rm p}$, see Fig. \ref{per1}b-c. The continualization of the particles of the set gives the smooth density function $\rho_{\rm c,p}({\bf r})$ and the periodic extension is
\begin{equation}
\rho_c({\bf r})=\sum_{n=-\infty}^{\infty}\rho_{\rm c,p}({\bf r}+n{\bf r}_{\rm p})
\: \: .
\end{equation}
The particles in $S_{\rm p}$ and the continuous function $\rho_{\rm c,p}({\bf r})$ generate the periodic behavior, but they are not restricted to a certain volume, as the volume of one period. 

As for two dimensional charge distributions with infinite size into the third dimension, the potential of periodic distributions is usually not limited to a finite range. Therefore the Poisson integral Eq. (\ref{Poisson_integral}) diverges for all observation points of interest! It is possible to avoid this and to achieve convergence, by normalizing the potential 
\begin{equation}
\tilde{\phi_c}({\bf r})= \phi_c({\bf r})-\phi_c({\bf r}_0)=
\frac{1}{4 \pi \varepsilon_0} \int dV' \times \rho({\bf r}') \left( \frac{1}{\| {\bf r}-{\bf r}'\|}
                                                                   -\frac{1}{\| {\bf r}_0-{\bf r}'\|}
                                                             \right)
\label{modified_Poisson_integral}
\end{equation}
to be zero at a certain point ${\bf r}_0$.
The gradient of $\tilde{\phi_c}$ is identical to that of $\phi_c$. In the following we skip the tilde
and write for the potential of the periodic distribution
\begin{eqnarray}
\phi_c({\bf r})&=&
\frac{1}{4 \pi \varepsilon_0} \int dV' \times \sum_{n=-\infty}^{\infty}\rho_{\rm p}({\bf r}'+n{\bf r}_{\rm p}) 
                                                             \left( \frac{1}{\| {\bf r}-{\bf r}'\|}
                                                                   -\frac{1}{\| {\bf r}_0-{\bf r}'\|}
                                                             \right)
 \: \: \nonumber \\
              &=&
\frac{1}{4 \pi \varepsilon_0} \int dV' \times \rho_{\rm p}({\bf r}') \sum_{n=-\infty}^{\infty} 
                                                             \left( \frac{1}{\| {\bf r}  +n{\bf r}_{\rm p}-{\bf r}'\|}
                                                                   -\frac{1}{\| {\bf r}_0+n{\bf r}_{\rm p}-{\bf r}'\|}
                                                             \right)
 \: \: . \nonumber
\end{eqnarray}
The summation term in the second integral can be interpreted as Green's function of a periodically repeated point charge.
The Green's function of one mesh cell follows from the integration
\begin{equation}
G_{\rm p}({\bf r}) = \int_{-h_x/2}^{h_x/2}\int_{-h_y/2}^{h_y/2}\int_{-h_z/2}^{h_z/2} dV' \times 
                                                             \sum_{n=-\infty}^{\infty} 
                                                             \left( \frac{1}{\| {\bf r}  +n{\bf r}_{\rm p}-{\bf r}'\|}
                                                                   -\frac{1}{\| {\bf r}_0+n{\bf r}_{\rm p}-{\bf r}'\|}
                                                             \right)
\: \: .  \nonumber
\end{equation}
This can be expressed by the non-periodic Green's function as
\begin{equation}
   G_{\rm p}({\bf r}) = \sum_{n=-\infty}^{\infty} G({\bf r}  +n{\bf r}_{\rm p}) - G({\bf r}_0  +n{\bf r}_{\rm p})
\: \: .
\end{equation}

To compute the potential at mesh points $\phi_{\rm m}(j,k,l)$, caused by all mesh charges, one has to
perform the summation
\begin{equation}
\phi_{\rm m}(j,k,l)=\frac{1}{4 \pi \varepsilon_0} \frac{1}{h_x h_y h_z} \sum_{j',k',l'}  q_{\rm c,p}(j',k',l') g_{\rm p}(j-j',k-k',l-l')
\: \: ,
\end{equation}
with $q_{\rm c,p}(j,k,l)$ the integrated charge of density $\rho_{\rm c,p}$ in mesh cell $(j,k,l)$ and $g_{\rm p}(j,k,l)$ the periodic Green's function at the mesh points. Therefore the field calculation for periodic source distributions can be done by the same method as for a non-periodic source, only the Green's function and the source distribution have to be replaced by the periodic extension $g_{\rm p}$ and the non-repeated representation $q_{\rm c,p}$.

\subsection{Numerical Calculation of Periodic Green's Function}

For the numerical evaluation we set ${\bf r}_0={\bf 0}$ and split $G_p$ into the terms
\begin{eqnarray}
G_1&=&\sum_{n=-M+1   }^{M-1 }\left( G({\bf r}+n{\bf r}_{\rm p} )- G(n{\bf r}_{\rm p})\right) \: \: ,\\
G_2&=&\sum_{n=-\infty}^{-M  }\left( G({\bf r}+n{\bf r}_{\rm p} )- G(n{\bf r}_{\rm p})\right) +
             \sum_{n=M>0    }^\infty\left( G({\bf r}+n{\bf r}_{\rm p} )- G(n{\bf r}_{\rm p})\right).
\end{eqnarray}
The finite sum is computed directly, using Eq.~(\ref{switched_greens_function}).
The number $M$ has to be chosen large enough, to fulfill the condition $\| {\bf r} \pm M {\bf r}_{\rm p}\|  > C_2 \max \{h_x, h_y, h_z \}$ for all points in the computation volume, compare Eq. \ref{criterion_for_far}.
In order to calculate the infinite sums, we use the asymptotic Eq.~(\ref{approximated_greens_function}) 
\begin{eqnarray}
G_2({\bf r}) &\approx& h_x h_y h_z  \left(  \sum_{n=-\infty}^{-M } \left( \frac{1}{\|{ \bf r}+n{\bf r}_{\rm p} \|}- \frac{1}{|n| r_{\rm p}} \right)
                                           +\sum_{n=M}^{\infty }   \left( \frac{1}{\|{ \bf r}+n{\bf r}_{\rm p} \|}- \frac{1}{|n| r_{\rm p}} \right) 
                                    \right) =\nonumber \\
                    &=& \frac{h_x h_y h_z}{r_{\rm p}} 
                    f\left(\frac{\bf{r}\cdot\bf{r}_{\rm p}}{r_{\rm p}^2} ,
                           \sqrt{\frac{r^2}{r_p^2}-\left( \frac{\bf{r}\cdot\bf{r}_{\rm p}}{r_{\rm p}^2} \right)^2},
                           M
                     \right)  ,
\end{eqnarray}
and the auxiliary function
\begin{eqnarray}
f(p,q,M>0)&=&\sum_{n=M}^\infty \left( \frac{1}{\sqrt{(n+p)^2+q^2}}+\frac{1}{\sqrt{(-n+p)^2+q^2}}-\frac{2}{n}\right)
\\
f(p,q,0)  &=& \frac{1}{\sqrt{p^2+q^2}}+f(p,q,1)
\: \: .
\end{eqnarray}

The auxiliary function $f(p,q,0)$ is periodic in $p$ and can be written as fourier series
\begin{equation}
 f(p,q,0)=f_{\rm f}(p,q) = const -2 \ln |q| + 4\sum_{n=1}^{\infty} K_0(2\pi n |q|) \cos(2\pi n p)
\: \: .
\end{equation}
The harmonic sum converges well for large values of $|q|$, due to the $\exp(-x)/\sqrt{x}$ behavior of the modified Bessel function for large arguments. Function values for positive $M$ are calculated by adding the finite sum $f(p,q,M)-f(p,q,0)$.

For small values of $q$ the auxiliary function is calculated with help of a  Taylor expansion
\begin{equation}
f(p,q,M>0) = f_t(p,q,M)+O\left(\left( p^2+q^2 \right)^5 M^{-10} \right)
\end{equation}
and the Taylor polynomial
\begin{eqnarray}
f_{\rm t}(p,q,M)&=& \left(2p^2-q^2\right ) S_{3,M} + \left(2p^6-6p^2q^2+\frac{3}{4}q^4 \right) S_{5,M} +
\nonumber \\    & &+\left(2p^6-15p^4q^2+\frac{45}{4}p^2q^4-\frac{5}{8}q^6 \right) S_{7,M} +
\nonumber \\    & &+\left(2p^8-28p^6q^2+\frac{105}{2}p^4q^4-\frac{35}{2}p^2q^6+\frac{35}{64}q^8 \right) S_{9,M}
\: \: ,
\end{eqnarray}
with $S_{k+1,M}=-\psi(k,M)/(k!)$ and $\psi$ the polygamma function. To reach the desired accuracy it can be necessary to split $f(p,q,M)$ into a finite sum $f(p,q,M)-f(p,q,M_f)$ and an infinite part $f(p,q,M_f)$ that is calculated by the Taylor approximation for a sufficiently high number $M_t$.

\begin{figure}
\includegraphics[width=1.00\textwidth]{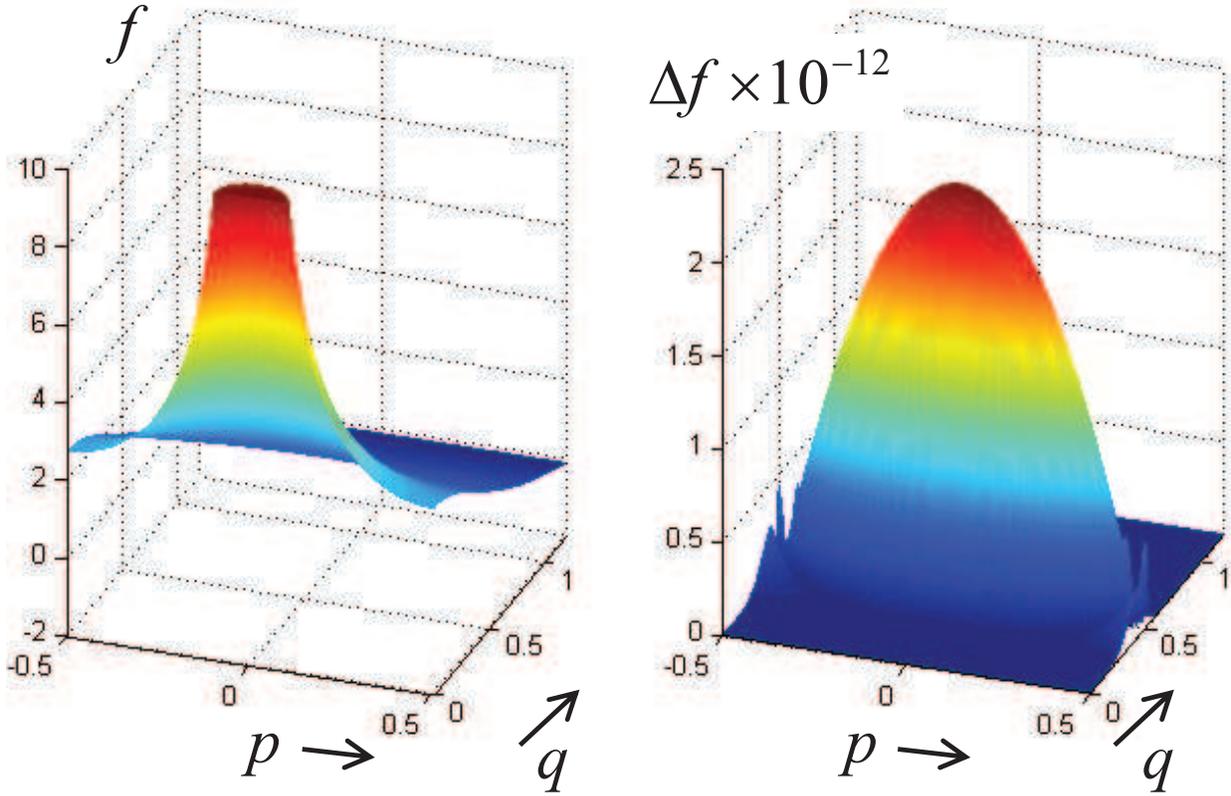} 
\caption{Function $f(p,q,0)$ and absolut error $\Delta f(p,q,0)$ of the numerical calculation. As $f$ is singular at the origin ($p=q=0$), it truncated for the plot to values below 10. The transition from the Taylor approximation (for small $|q|$) to the Fourier series is at $|q|=0.5$. The Taylor approximation is used for the upper part of the sum from $M_t=16$ to infinity. The harmonic expansion is truncated after 8 terms.}
\label{SuF}
\end{figure}

The numerically calculated function $f(p,q,0)$ and the deviation $\Delta f(p,q,0)=|f(p,q,0)-f_{\rm hp}(p,q,0)|$ from a ``high precession'' result are plotted in Fig. \ref{SuF} for $p \in [-0.5,0.5]$ and $q \in [0,1.2]$. The ``high precession'' result $f_{\rm hp}(p,q,0)$ has been calculated to numerical accuracy by using a Fourier series to sufficiently high harmonics or by a sufficiently high $M_t$ for the Taylor expansion. The highest values of $\Delta f$ appear at $q \simeq 0.5$ where the type of expansion is switched. The Fourier expansion is truncated after 8 terms, the Taylor expansion uses $M_t=16$.

\subsection{Mesh-Periodicity and Symmetry}

\begin{figure}
\includegraphics[width=0.75\textwidth]{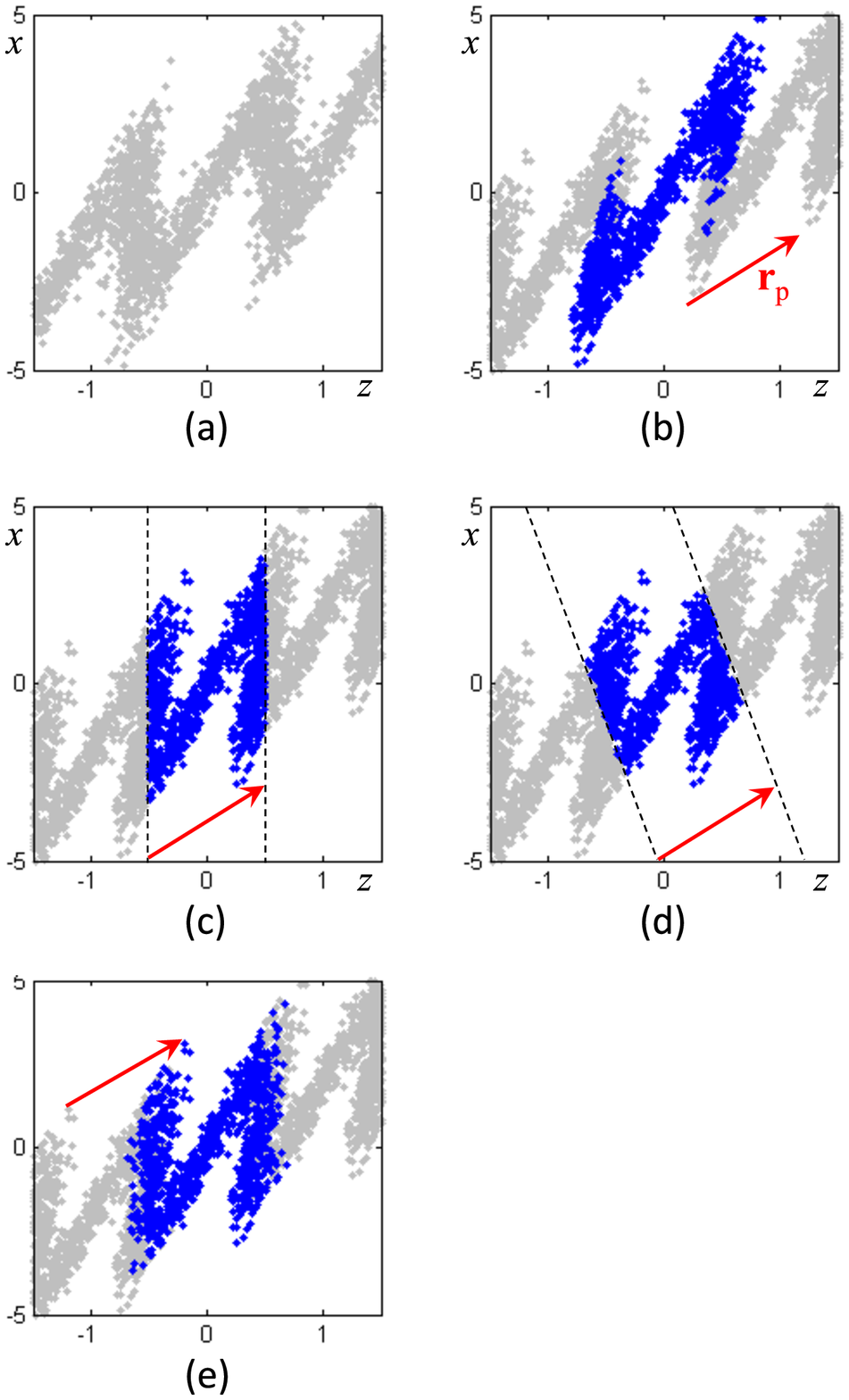} 
\caption{Periodic particle distributions and periodic shift vector ${\bf r}_{\rm p}$: (a) pseudo-periodic random distribution with periodic behavior; (b) generating set $S_{\rm p}$ and periodically repeated particles in blue resp. gray; (c) equivalent generating set in slice volume between two planes shifted by ${\bf r}_{\rm p}$; (d) other equivalent set in different volume slice; (e) equivalent generating set without sharp truncation.}
\label{per1}
\end{figure}

\begin{figure}
\includegraphics[width=0.90\textwidth]{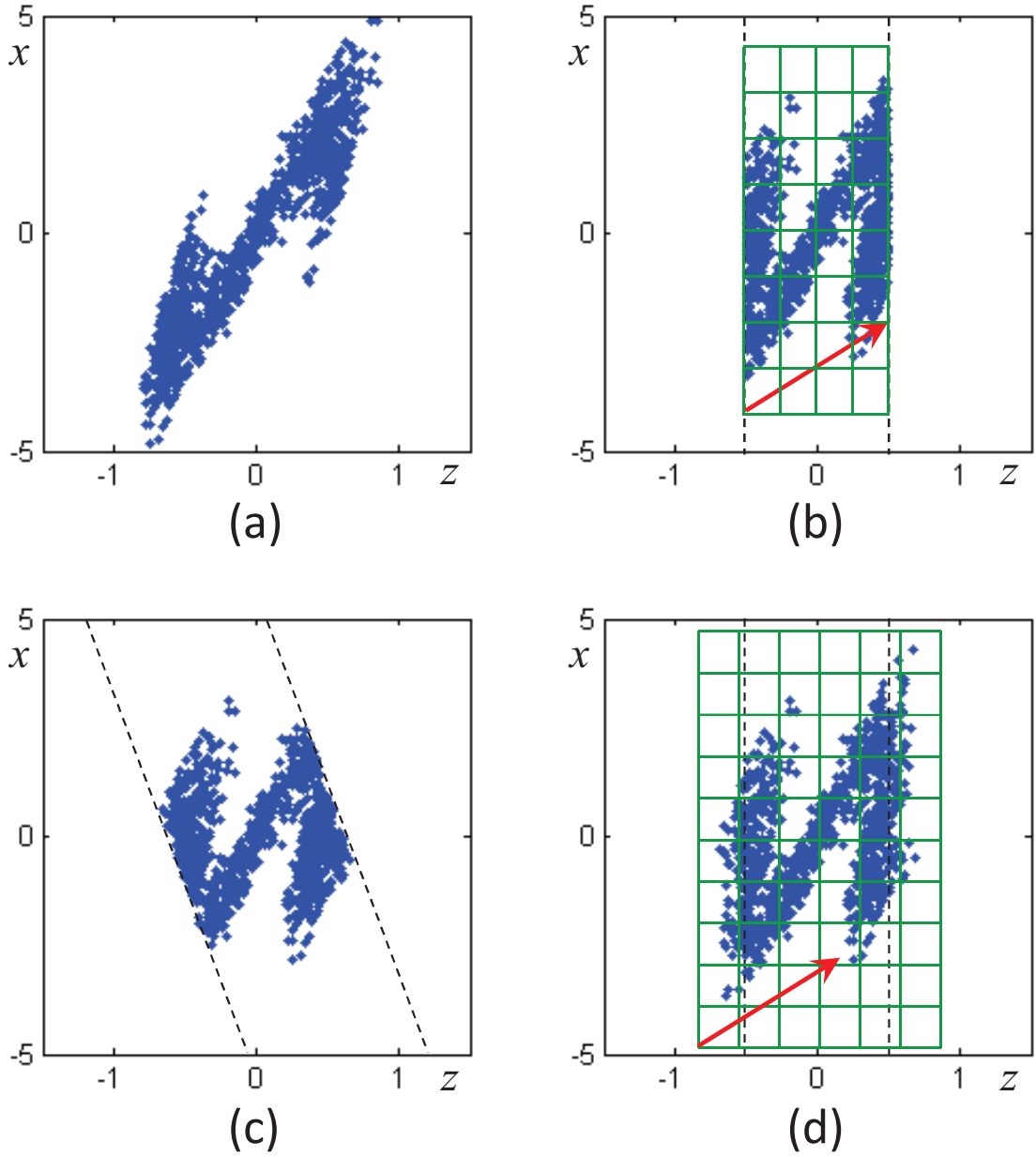} 
\caption{Equivalent generating sets $S_{\rm p}$ for the same periodic distribution: (a) volume is not limited to a one-period-slice; (b) particles in one-period-slice and mesh that supports periodicity; (c) particles in equivalent one-period-slice; (d) particles without sharp truncation and mesh that does not support periodicity.}
\label{per2}
\end{figure}

It is obvious from Fig. \ref{per1} that the choice of $S_{\rm p}$ is not unique. Nevertheless the convolution of the continualizated density with the periodic Green's function might result in the same values $\phi_{\rm m}(j,k,l)$ of the potential on the mesh. This happens  if the mesh supports the periodicity, which means the vector ${\bf r}_{\rm p}$ coincides with a mesh vector $j h_x {\bf e}_{\perp 1}+k h_y {\bf e}_{\perp 2}+l h_z {\bf e}_{\rm c}$ as in Fig. \ref{per2}b.

For practical simulations it is desirable to use a mesh that supports periodicity, and to shift the particles of the original set by a multiple of ${\bf r}_{\rm p}$ so that they fill a one-period-slice as in Fig. \ref{per1}b and c. It stands to reason, to align the slice with respect to the mesh as in Fig. \ref{per1}b. A small volume of filled mesh cells is preferable.

Meshes that do not support periodicity can be favorable to avoid extreme aspect ratios of mesh cells, as it would happen for supported periodicity if the ratio of non-zero components of ${\bf r}_{\rm p}$ is extreme. Such a mesh, as in Fig. \ref{per2}d, gives reasonable results supposed the generating set $S_{\rm p}$ is sufficiently smooth. (Sudden changes of density have to be resolved by the discretization.) The potential has to be computed in a volume that
overlaps partially with the shifted volume. In overlapping regions the potential is not quite identical, due to non-periodic discretization errors. Therefore an appropriate interpolation between potentials in overlapping regions (while entering one region and leaving the other) helps to avoid discontinuous potentials and infinite fields.

The periodical Green's function has point symmetry to the origin $G_{\rm p}(j,k,l)=G_{\rm p}(-j,-k,-l)$. It is mirror symmetric $G_{\rm p}(j,k,l)=G_{\rm p}(-j,k,l)$ if ${\bf r}_{\rm p} \cdot {\bf e}_{\perp 1}=0$, or $G_{\rm p}(j,k,l)=G_{\rm p}(j,-k,l)$ if ${\bf r}_{\rm p} \cdot {\bf e}_{\perp 2}=0$, or $G_{\rm p}(j,k,l)=G_{\rm p}(j,k,-l)$ if ${\bf r}_{\rm p} \cdot {\bf e}_{\rm c}=0$.

\section{Applications}

\subsection{FLASH Seeding Section, with 800 nm Energy Modulation of a 1.5 kA Bunch}
\begin{figure}
\includegraphics[width=1.0\textwidth]{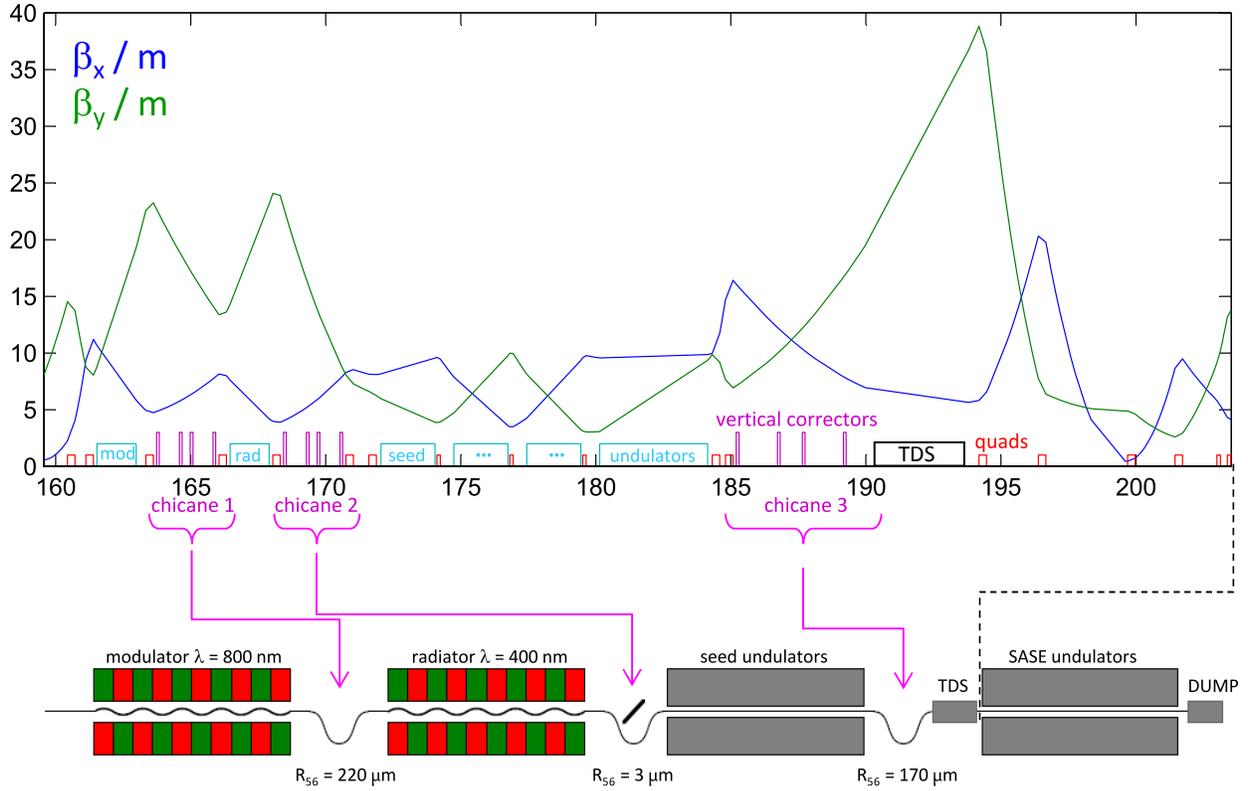} 
\caption{Seeding section of FLASH: optical functions and layout.}
\label{flash_seeding_section}
\end{figure}

In the seeding section of the FLASH free electron laser
\cite{FLASH, Hacker, SASE_Sup}
an electron bunch is energy modulated by an external laser and send through three chicanes and several undulators as it can be seen in Fig.~\ref{flash_seeding_section}. To investigate microbunching effects the radiator undulator (after chicane 1) and the sFLASH undulators (after chicane 2) have been switched off. The simulation starts directly after the modulator (longitudinal position 162 m) and ends 10 m  behind after the transverse deflecting structure TDS (at longitudinal position 204.2 m).

The initial beam (after the modulator) has the following properties: energy ${\cal E}_0$ = 585 MeV, charge 0.3 nC, normalized emittance 1.5 $\mu$m, peak current 1.5 kA, uncorrelated energy spread $\sigma_{\cal E}$ = 150 keV and Gaussian longitudinal and transverse profiles. Therefore, the rms length is 80 fs. The beam has been energy modulated in the modulator undulator by a laser with wavelength $\lambda$ = 800 nm. The duration of the laser pulse of minimal 30 fs (FWHM) is short compared to bunch length, but long enough to permit pseudo periodic behavior. As the optical beam in the undulator is much wider than the particle beam, the energy modulation amplitude ${\cal \hat E}$ (of 250 keV) is nearly transverse offset independent.

\begin{figure}
\includegraphics[width=0.44\textwidth]{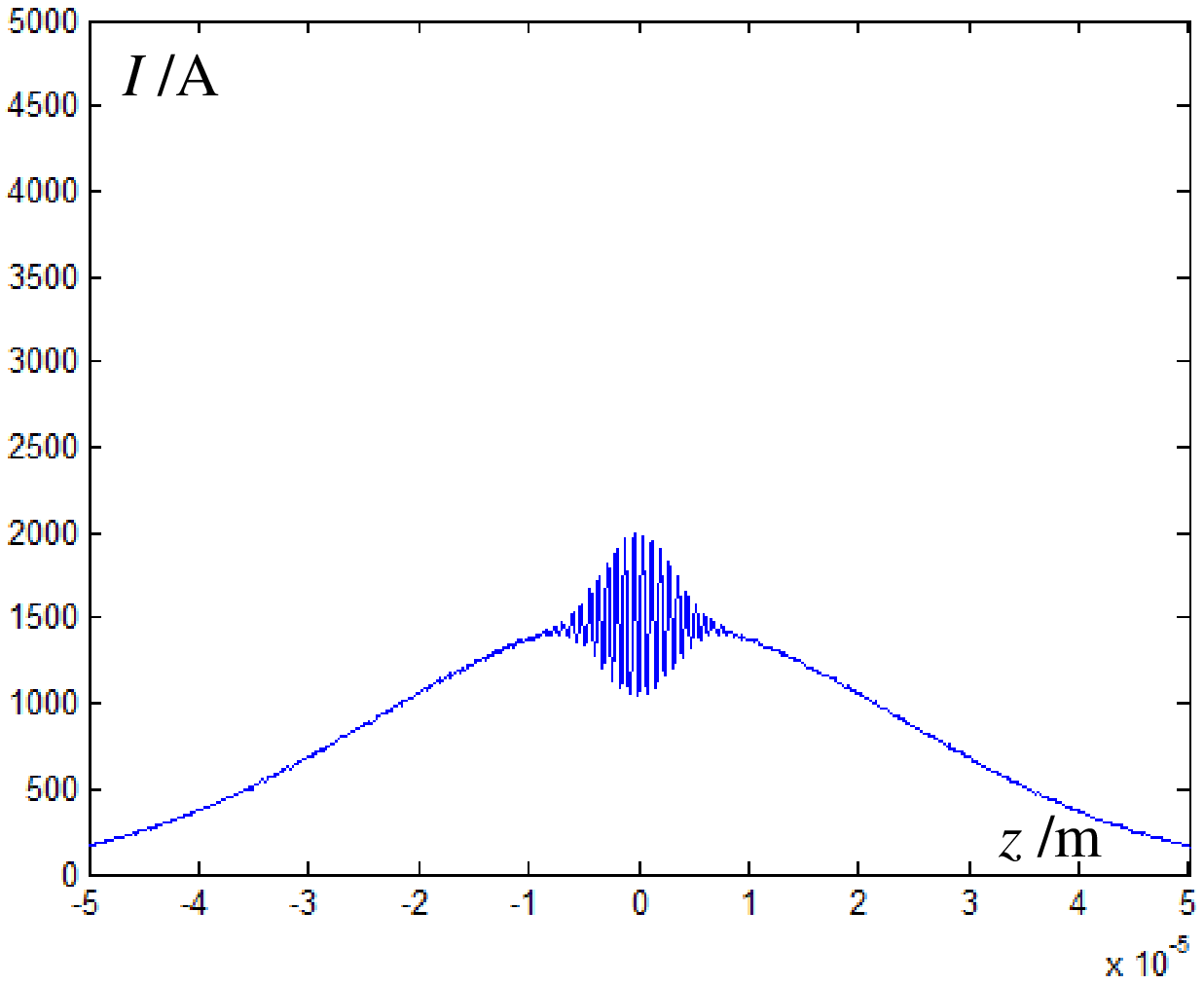} 
\includegraphics[width=0.44\textwidth]{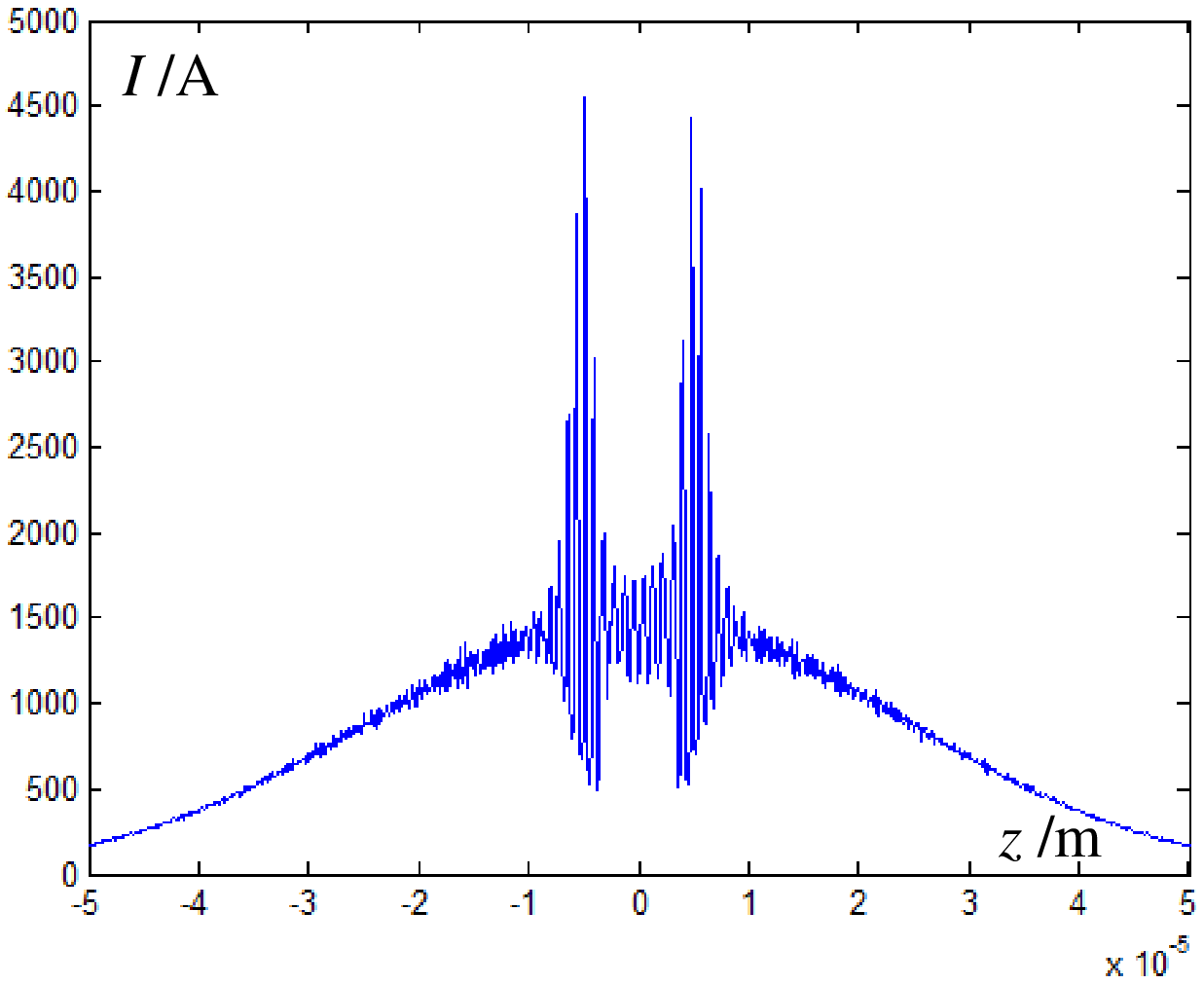} 
\\ \hspace{0.3cm} (a) \hspace{6.7cm} (b)
\caption{Application A: current (a) before and (b) after the third chicane, for 30 fs (FWHM) laser modulation, calculated with 20E6 macro-particles.}
\label{Cur_250keV_before_after_CH3}
\end{figure}

\begin{figure}
\includegraphics[width=0.44\textwidth]{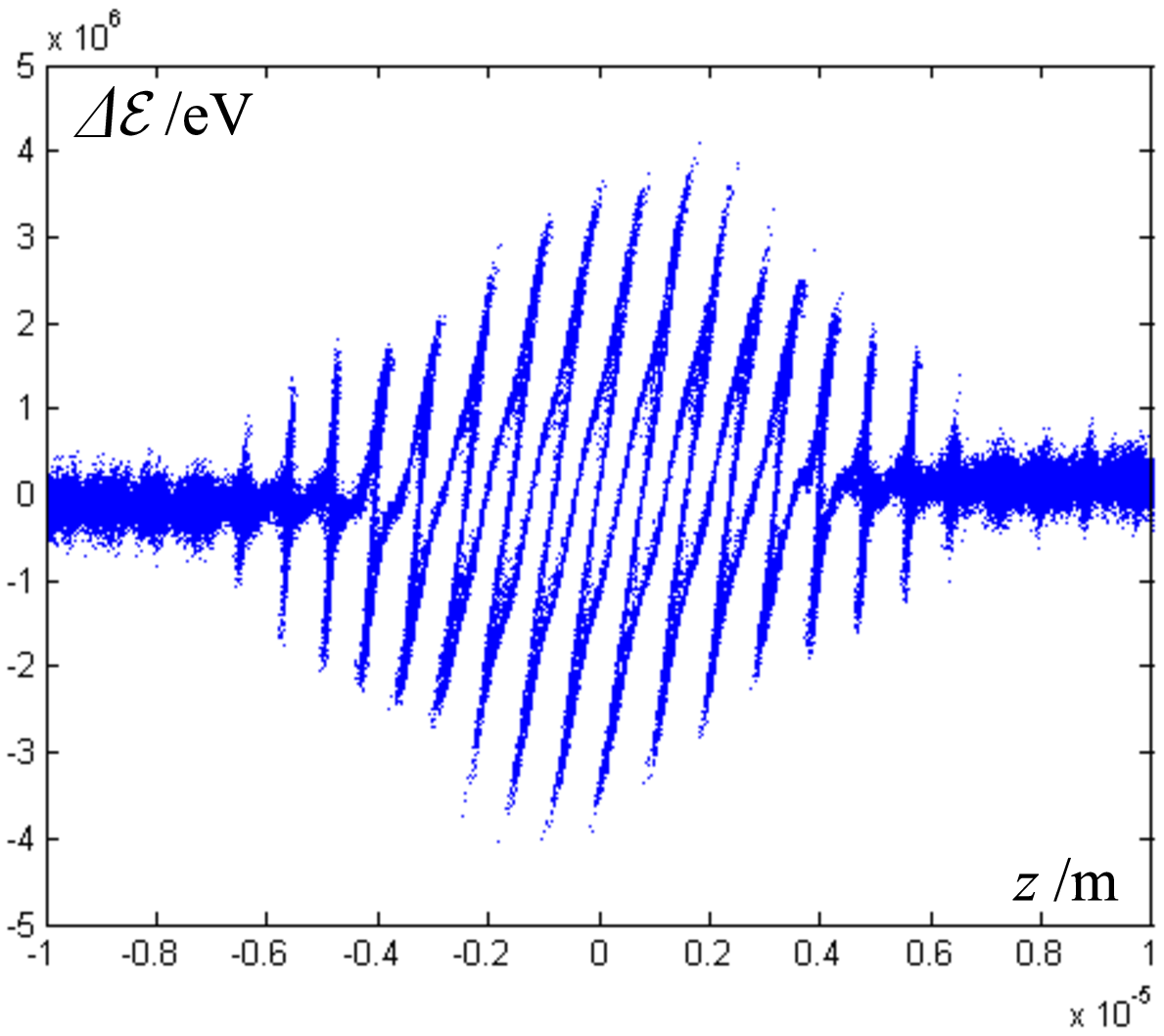} 
\includegraphics[width=0.44\textwidth]{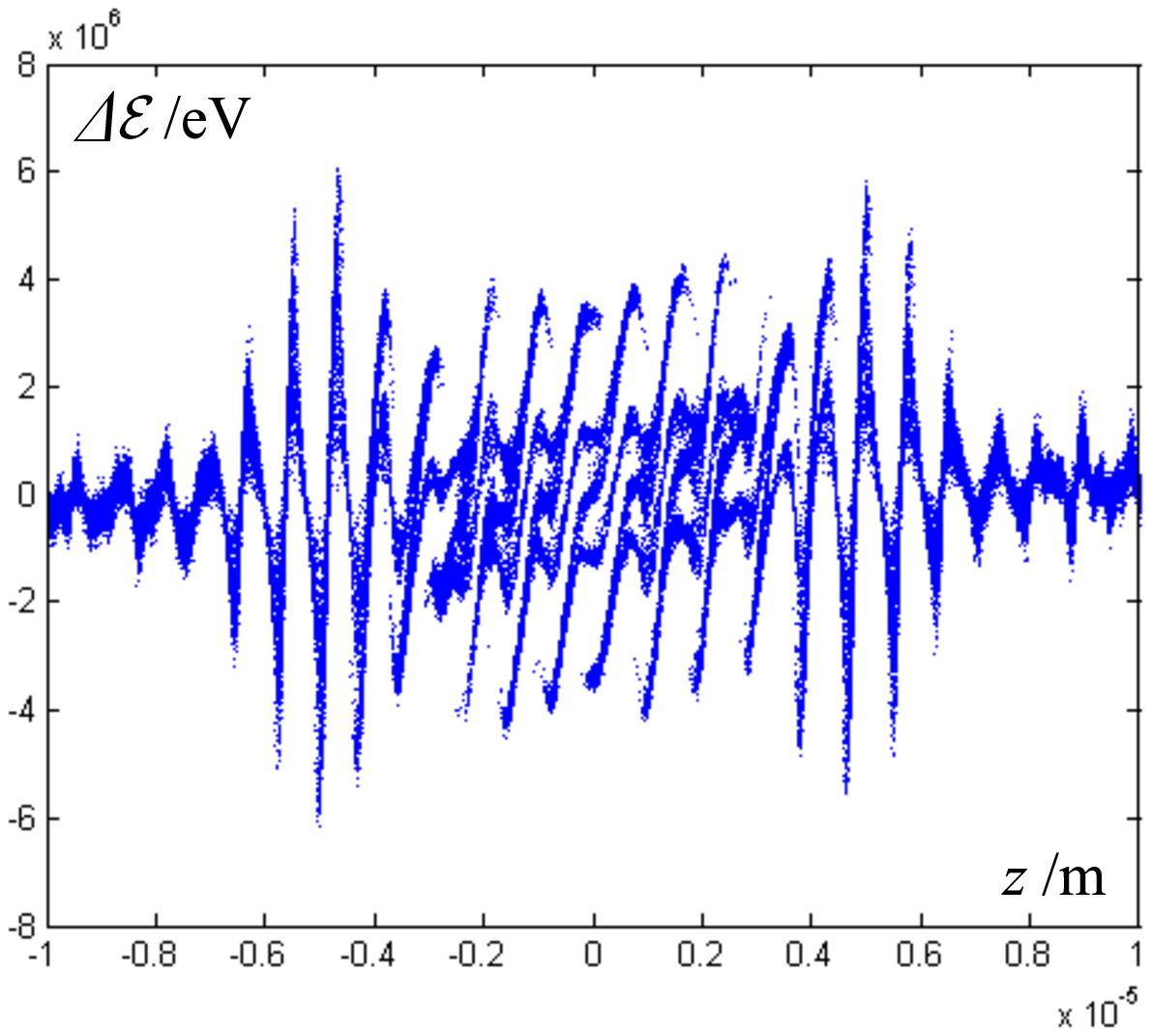} 
\\ \hspace{0.3cm} (a) \hspace{6.7cm} (b)
\caption{Application A: longitudinal phase space (a) direct after chicane 3 and (b) 15 m behind, for 30 fs (FWHM) laser modulation, calculated with 1E6 macro-particles, by Astra.}
\label{EvZ_250keV_189p4_204p2_Astra_xyz2}
\end{figure}

Effects in drift-quadrupole-lattices are the modification of particle energies, the modification of transverse optics and plasma oscillations. For the given parameters, the energy effect is strongest, transverse defocussing is weak and the length for a full plasma oscillation, in longitudinal phase space, is about $L_{\rm p}$ = 120 m
\cite{SPIE_Prag}.
Therefore the longitudinal phase space distribution is nearly not altered on the short distance to the first chicane. The longitudinal dispersion of the first chicane $R_{56}=220 \mu$m couples relative energy deviations $\delta_{\cal E}=({\cal E}-{\cal E}_0)/{\cal E}_0$ to longitudinal displacements $\delta z = R_{56}  \delta_{\cal E}$, and creates density modulation due to the initial energy modulation. The effect of the second chicane is weak due to its small dispersion $R_{56}=3 \mu$m. More important is the interchange between density and energy modulation, as it is typical for plasma oscillations, on the distance between chicane 1 and 3 of about $L_{\rm p}/6$. The bunch current with its modulation is shown in Fig~\ref{Cur_250keV_before_after_CH3}a. Up to this position, the energy modulation has grown by more than a factor of ten to an amplitude of about 4 MeV. This amplification is beyond the linear regime and this will be exacerbated in the last chicane with $R_{56}=170 \mu$m. This longitudinal dispersion is sufficient to shear particles, that were originally (during modulation) in one period, over several periods as it can be seen within the longitudinal phase space in Fig.~\ref{EvZ_250keV_189p4_204p2_Astra_xyz2}a. At rollover ``points'' in longitudinal phase space, where the band with particles is folded back, current spikes appear as can be seen in Fig.~\ref{Cur_250keV_before_after_CH3}b. These spikes cause strong space charge fields and lead to a complicated energy pattern after a further drift of about 15 m, see Figs.~\ref{EvZ_250keV_189p4_204p2_Astra_xyz2}b.

\begin{figure}
\includegraphics[width=0.44\textwidth]{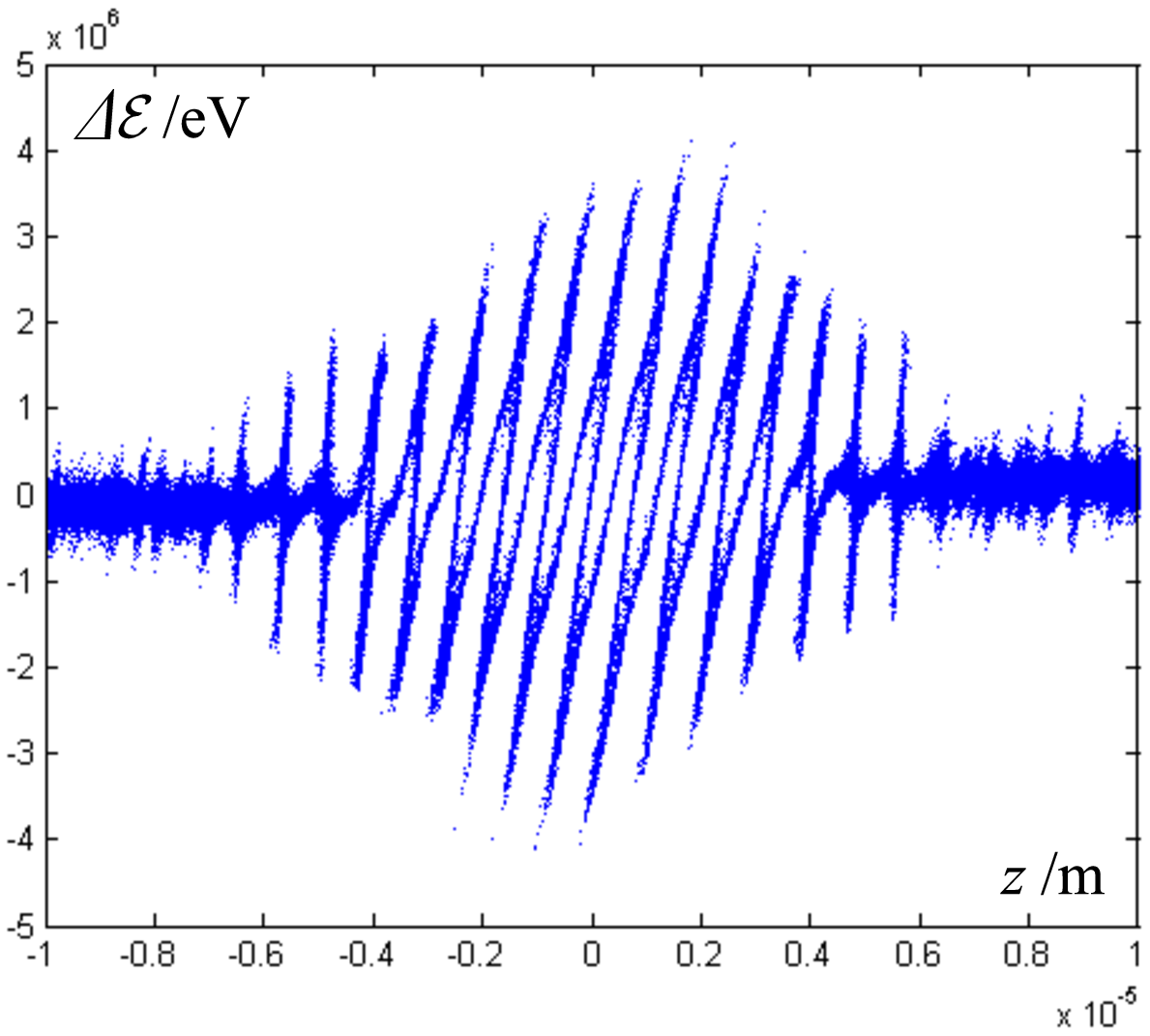} 
\includegraphics[width=0.44\textwidth]{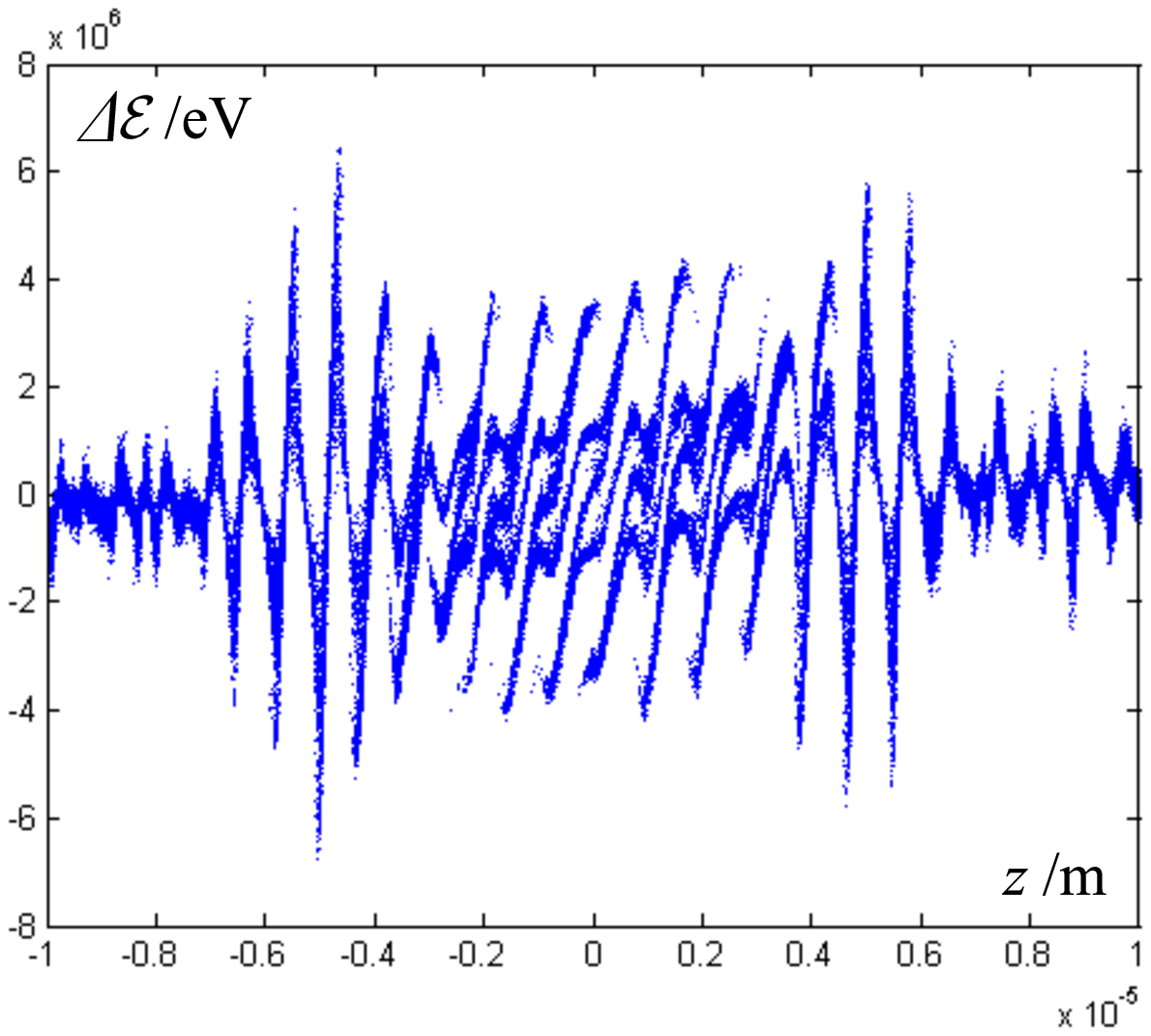} 
\\ \hspace{0.3cm} (a) \hspace{6.7cm} (b)
\caption{Application A: longitudinal phase space (a) direct after chicane 3 and (b) 15 m behind, for 30 fs (FWHM) laser modulation, calculated with 1E6 macro-particles, by QField.}
\label{EvZ_250keV_189p4_204p2_QField_xyz}
\end{figure}

Fig.~\ref{EvZ_250keV_189p4_204p2_Astra_xyz2} has been calculated by the established space charge code Astra
\cite{Floettmann}
with a three dimensional solver based on the particle mesh method. For comparison we did non-periodic and periodic simulations with our implementation, called QField.
The results of full-bunch simulations with 1E6 macro particles with Astra and QField-none-periodic can be seen in Figs.~\ref{EvZ_250keV_189p4_204p2_Astra_xyz2} and \ref{EvZ_250keV_189p4_204p2_QField_xyz}, showing the longitudinal phase space after chicane 3 and 15 m downstream. The results are in good agreement.

\begin{figure}
\includegraphics[width=0.66\textwidth]{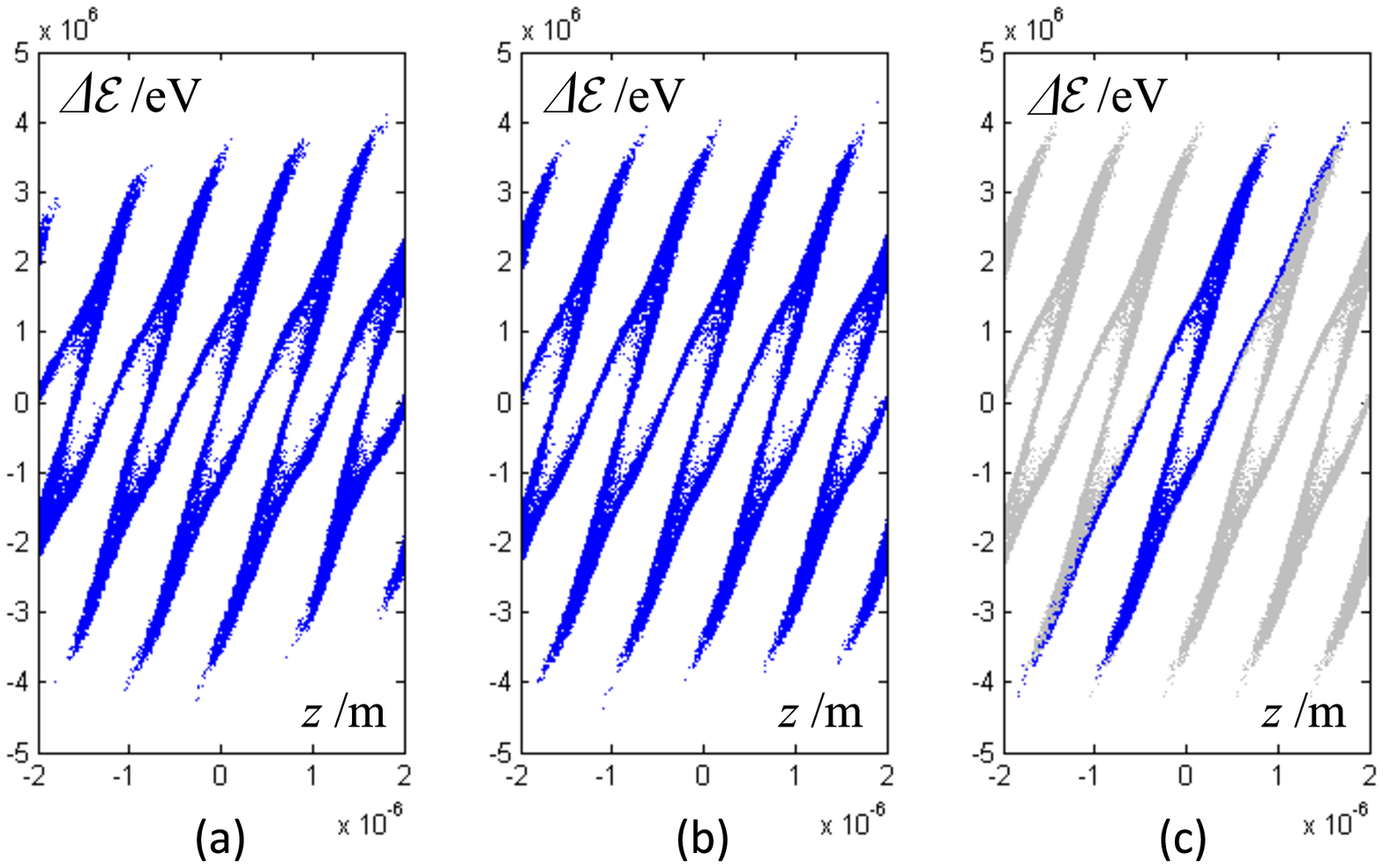} 
\caption{Application A: longitudinal phase space direct after chicane 3. (a) Calculated with QField-non-periodic, 1E6 macro-particles, 30 fs (FWHM) laser modulation, (b) calculated with QField-non-periodic, 20E6 particles, 60 fs (FWHM) laser modulation, (c) calculated with QField-periodic.}
\label{fig_189p4_1E6_20E6_per}
\end{figure}

\begin{figure}
\includegraphics[width=0.66\textwidth]{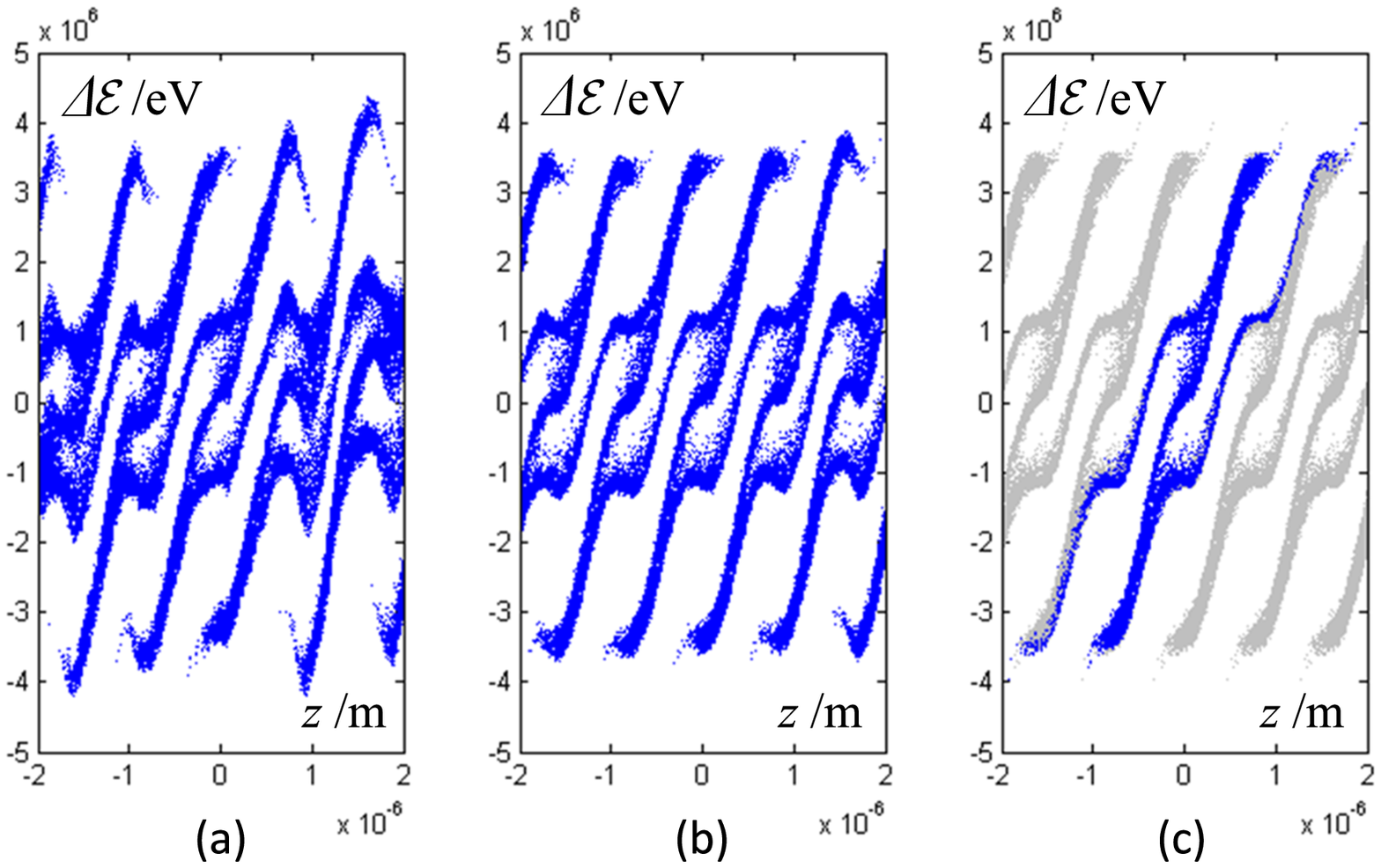} 
\caption{Application A: longitudinal phase space 15 m behind chicane 3. (a) Calculated with QField-non-periodic, 1E6 macro-particles, 30 fs (FWHM) laser modulation, (b) calculated with QField-non-periodic, 20E6 particles, 60 fs (FWHM) laser modulation, (c) calculated with QField-periodic.}
\label{fig_204p2_1E6_20E6_per}
\end{figure}

To verify the periodic approach, we did a high resolution full-bunch simulation with 20E6 macro particles with QField-none-periodic, and a periodic simulation with same macro particle density in only one period. The full-bunch simulations was done with longer laser modulation (60 fs, FWHM) to increase the length of the pseudo periodic range. Figs.~\ref{fig_189p4_1E6_20E6_per} and \ref{fig_204p2_1E6_20E6_per} show a short part of the longitudinal phase space around the middle of the bunch, directly after the last chicane and 15 m downstream. Particles in gray are periodic repetitions of the simulated particles that are plotted in in blue. Not all simulated particles in the observation range are plotted to avoid large over-painted areas. (The number of plotted particles is equivalent for the different plots.) In the sub-panels \ref{fig_189p4_1E6_20E6_per}a, \ref{fig_204p2_1E6_20E6_per}a for the full-bunch, the non-periodic behavior due to finite modulation length is clearly observable. The sub-panels \ref{fig_189p4_1E6_20E6_per}b, \ref{fig_204p2_1E6_20E6_per}b for the full-bunch simulation with longer laser modulation show a nice quasi-periodic pattern that is in good agreement with the periodic results in Figs.~\ref{fig_189p4_1E6_20E6_per}c, \ref{fig_204p2_1E6_20E6_per}c.

\subsection{FLASH Seeding Section, with 266 nm Energy Modulation of a 45 A Bunch}
This example compares the three-dimensional periodic model with a one-dimensional impedance model, for a periodic distribution that is modulated and compressed to achieve current spikes of maximal amplitude. Significant self effects happen on the drift-quadrupole-lattice after the four magnet compressor chicane. The one-dimensional impedance model averages the longitudinal field of a round Gaussian beam versus the transverse offset, as derived in
\cite{Saldin-1d-impedance}
, see also
\cite{SPIE_Prag}
. A periodic version of the one dimensional model was used for the comparison. (The periodic repetition can be realized either by a summation of the spatially shifted field, or by a summation of the harmonics of the impedance in frequency domain.)

\begin{figure}
\includegraphics[width=0.39\textwidth]{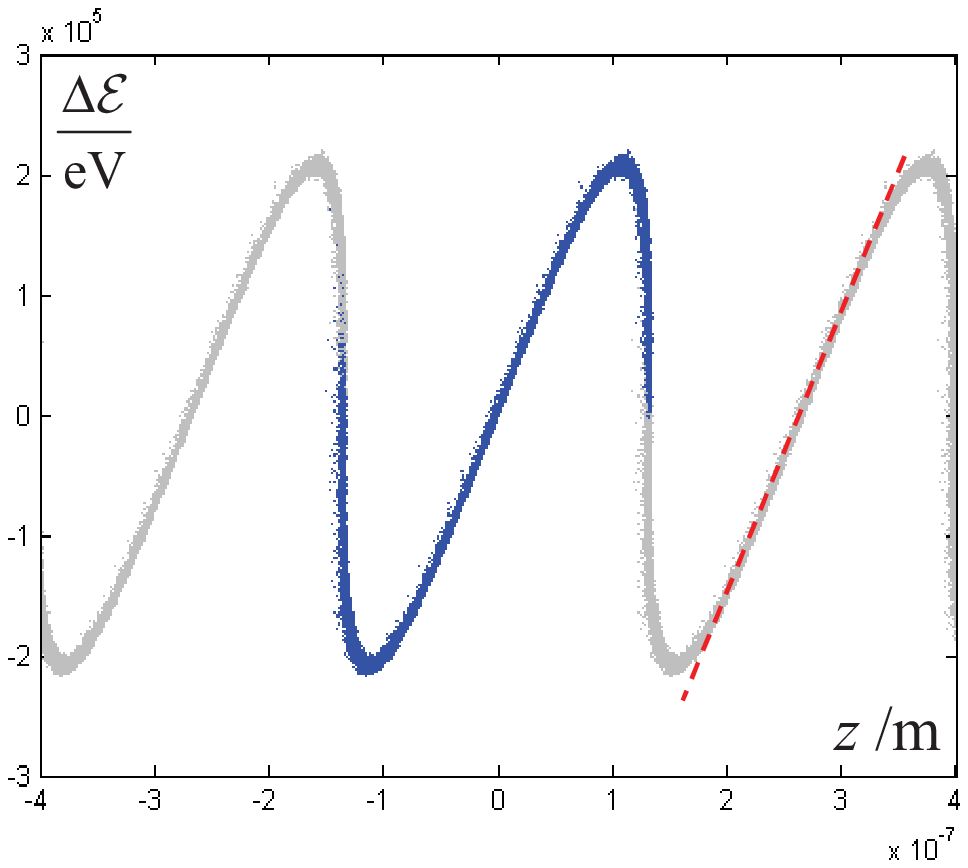} 
\includegraphics[width=0.47\textwidth]{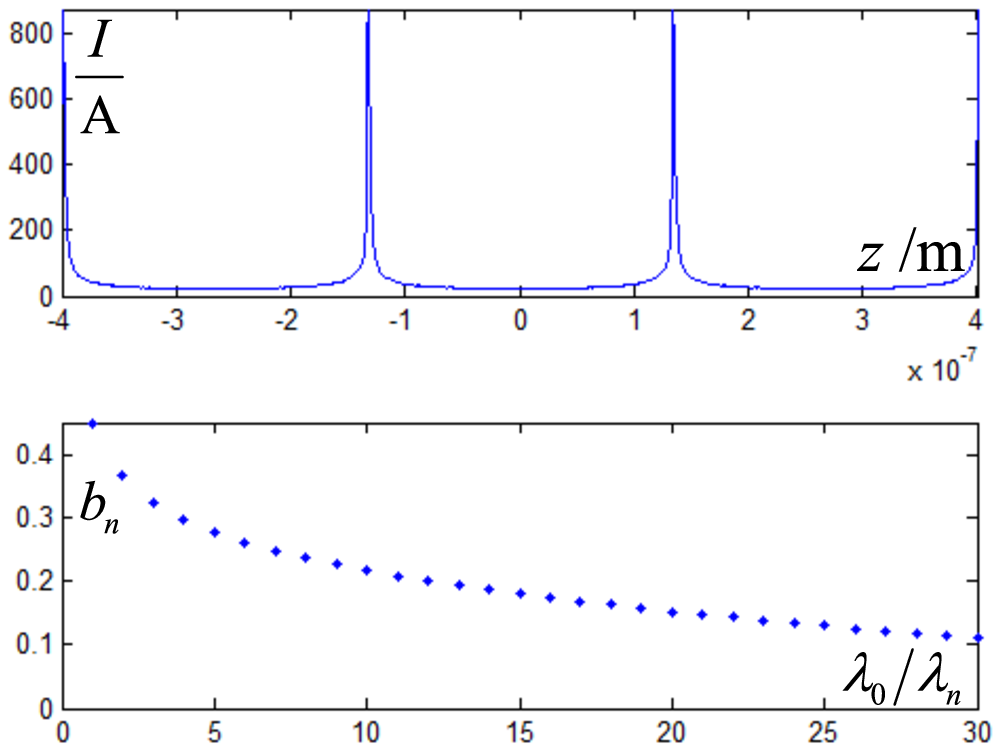} 
\caption{Application B: longitudinal phase space, current and bunching after 1st chicane.}
\label{EX2_165p92}
\end{figure}

The setup is the same as before, but with different parameters and only one active chicane: beam energy ${\cal E}_0$ = 700 MeV, bunch charge 0.3 nC, normalized emittance 1.5 $\mu$m, peak current 45 A, uncorrelated energy spread $\sigma_{\cal E}$ = 3keV and Gaussian profiles. The particle beam is energy modulated by a laser of wavelength $\lambda$ = 267 nm with the modulation amplitude ${\cal \hat E}$ = 0.2 MeV. The dispersion $R_{56} \approx 145 \mu$m of the first chicane is adjusted for maximum bunching and maximum current spikes as it can be seen Fig.~\ref{EX2_165p92}. The second and third chicane are switched off.

Directly before the first chicane, the phase space distribution is almost not changed by self effects, due to the small beam current, the high particle energy and the short travelled distance to this position. In contrast to the previous example, the particles are tracked through the magnetic fields of the chicane, but the effect of one- respectively three-dimensional self fields are so small, that Fig.~\ref{EX2_165p92} looks identical for both cases, and it is not to distinguish from the corresponding figure for a calculation without self-fields. The sharpness of the current spikes is not only determined by uncorrelated energy spread and longitudinal dispersion, but also by transverse emittance and second order coupling to the longitudinal direction. The current spikes are very short and the bunching factor
$b_n=\left| \sum{\exp(i n 2\pi z_\nu/\lambda)} \right|/N$
is above ten percent even for spectral lines beyond the 30th harmonic. The required resolution of the longitudinal mesh has to be much better than than $\lambda/30$. It is set to 0.6 nm, which is 450 meshlines per period of the fundamental modulation and thus below the shortest possible rms length of $R_{56} \sigma_{\cal E}/{\cal E}_0$ that would appear if the particles are ideally (linear chirp, no non-linear effects) compressed to minimal length. There are about 220000 electrons per period which are directly simulated. A macro particle number equal (or higher) than the number of electrons can be easily handled with the periodical approach. 

\begin{figure}
\includegraphics[width=0.39\textwidth]{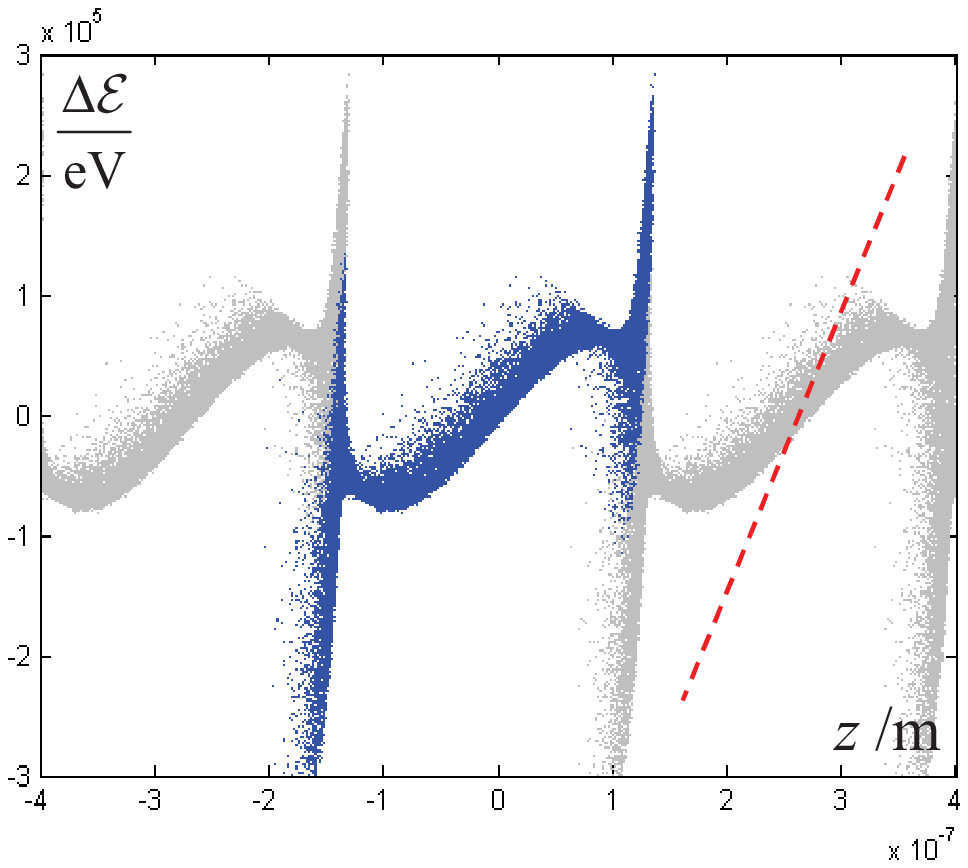} 
\includegraphics[width=0.47\textwidth]{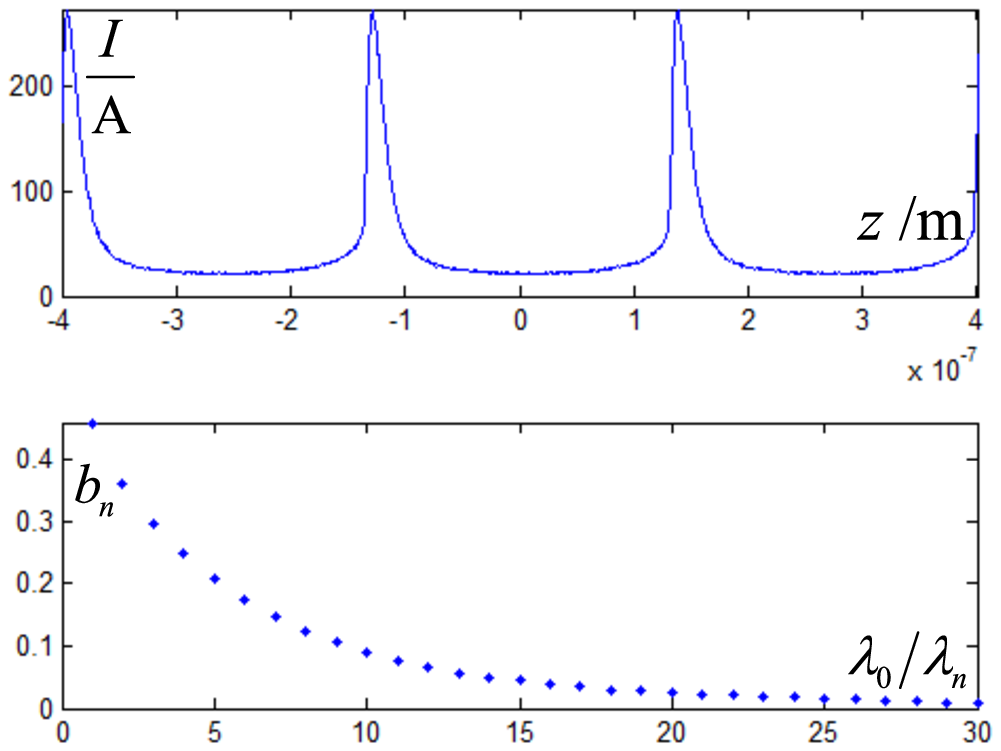} 
\caption{Application B: longitudinal phase space, current and bunching after 3rd chicane, calculated with one dimensional impedance. Chicanes 2 and 3 are off.}
\label{EX2_189p30_1d}
\end{figure}

\begin{figure}
\includegraphics[width=0.39\textwidth]{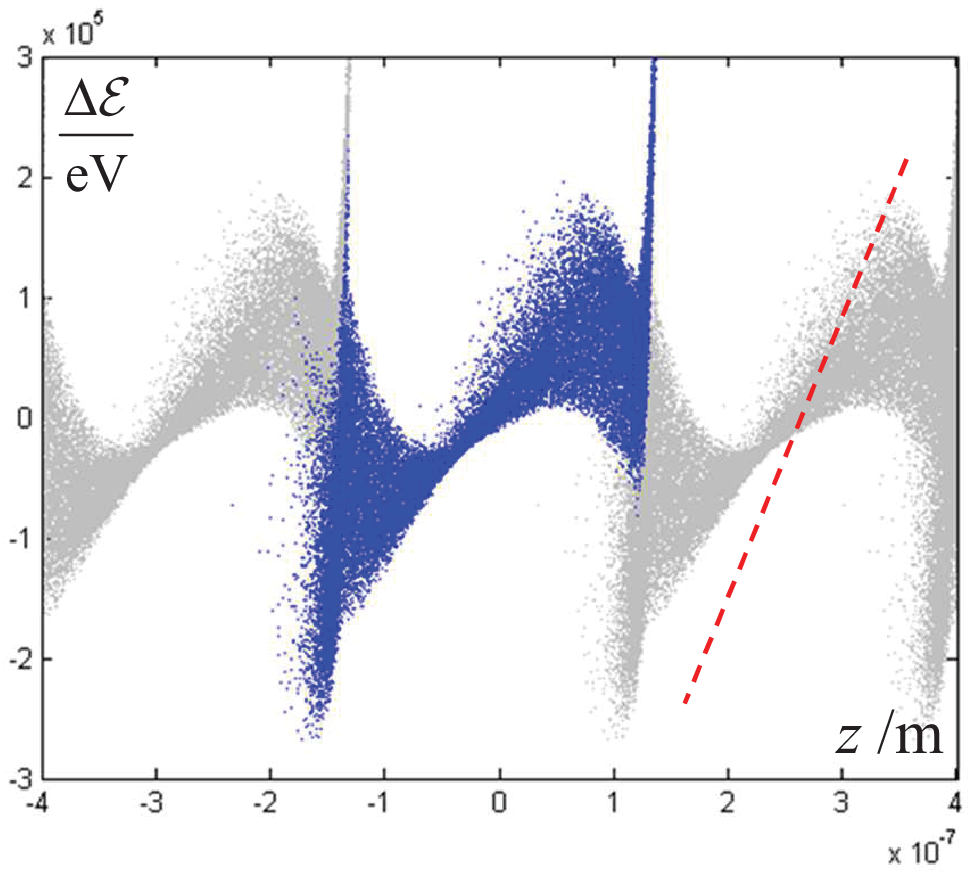} 
\includegraphics[width=0.47\textwidth]{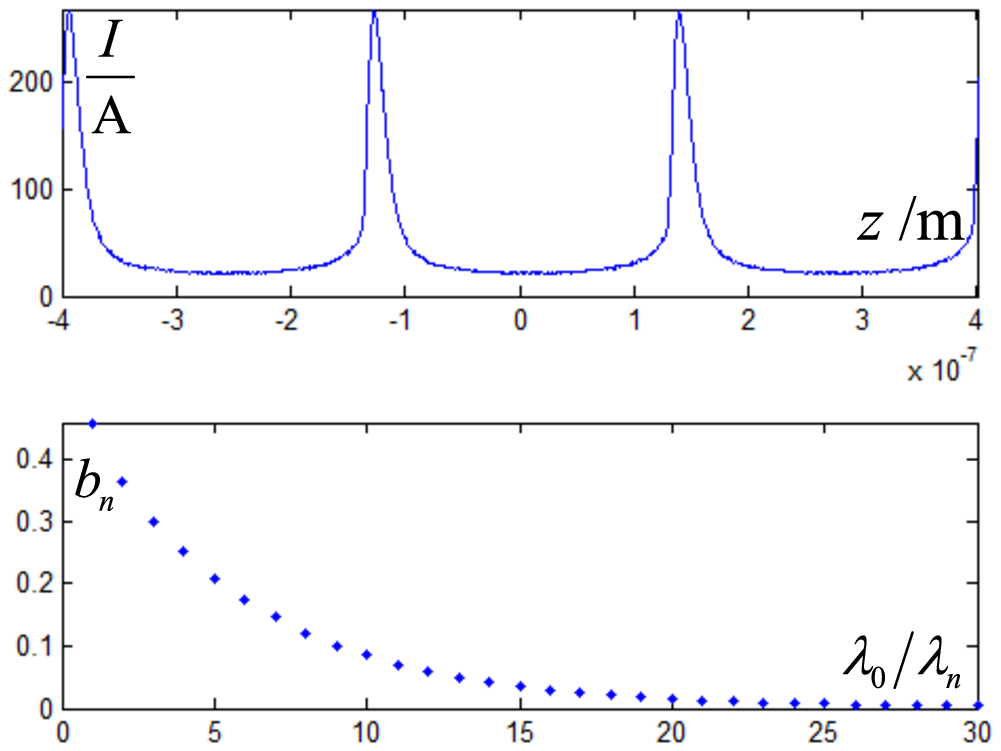} 
\caption{Application B: longitudinal phase space, current and bunching after 3rd chicane, calculated with three dimensional impedance. Chicanes 2 and 3 are off.}
\label{EX2_189p30_3d}
\end{figure}

The particle distribution with sharp density spikes is tracked about 23 m through quadrupoles and drifts to the exit of the (inactive) third chicane. Figs.~\ref{EX2_189p30_1d}, \ref{EX2_189p30_3d} show the longitudinal phase space, the bunch current and the bunching for the one- and three-dimensional field model. The current spikes are diverged and the high harmonic content is significantly reduced, in good agreement of both models. It can be seen that the slope of the initial saw-tooth-like modulation has been reduced, similar to the reduction of the energy modulation during a plasma oscillation, after a length that is short compared to the period length. In the one-dimensional model all particles in one slice feel the same longitudinal field. Therefore the only mechanism to increase the slice energy spread is individual longitudinal particle motion. The longitudinal field in the spikes is strongest and causes an energy modulation comparable to the initial laser modulation and a large energy spread. But the space charge distribution significantly violates the requirements for the one dimensional approach of a long bunch: $\gamma \sigma_z \gg \sigma_{\perp}$. The length of the current spike of few nano-meters times $\gamma$ is not large compared to the typical transverse dimension of about 100 $\mu$m (= emittance $\times \beta$, see Fig.~\ref{flash_seeding_section} with the beta function). The vectorial field of the three dimensional model depends on the transverse offset and is integrated along betatron trajectories. (For the other extreme of flat bunches $\gamma \sigma_z \ll \sigma_{\perp}$ the longitudinal field is proportional to the transverse density, as $\partial E_z/\partial z \approx {\rm div} {\bf E} = \rho/\varepsilon_0$.) This increases the slice energy spread as it can be seen in Fig.~\ref{EX2_189p30_3d}.

\subsection{Parasitic Heating after LCLS Laser Heater}
\begin{figure}
\includegraphics[width=0.66\textwidth]{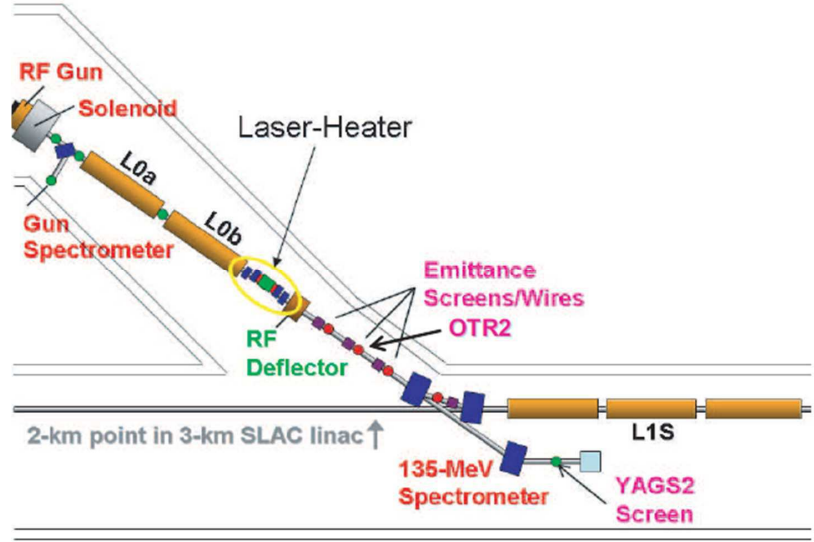} 
\caption{Application C: the LCLS injector layout showing laser heater, transverse RF deflector OTR/YAG screens wire scanners, and spectrometers, from \cite{Huang LH}}
\label{LCLS_LH}
\end{figure}

\begin{figure}
\includegraphics[width=0.80\textwidth]{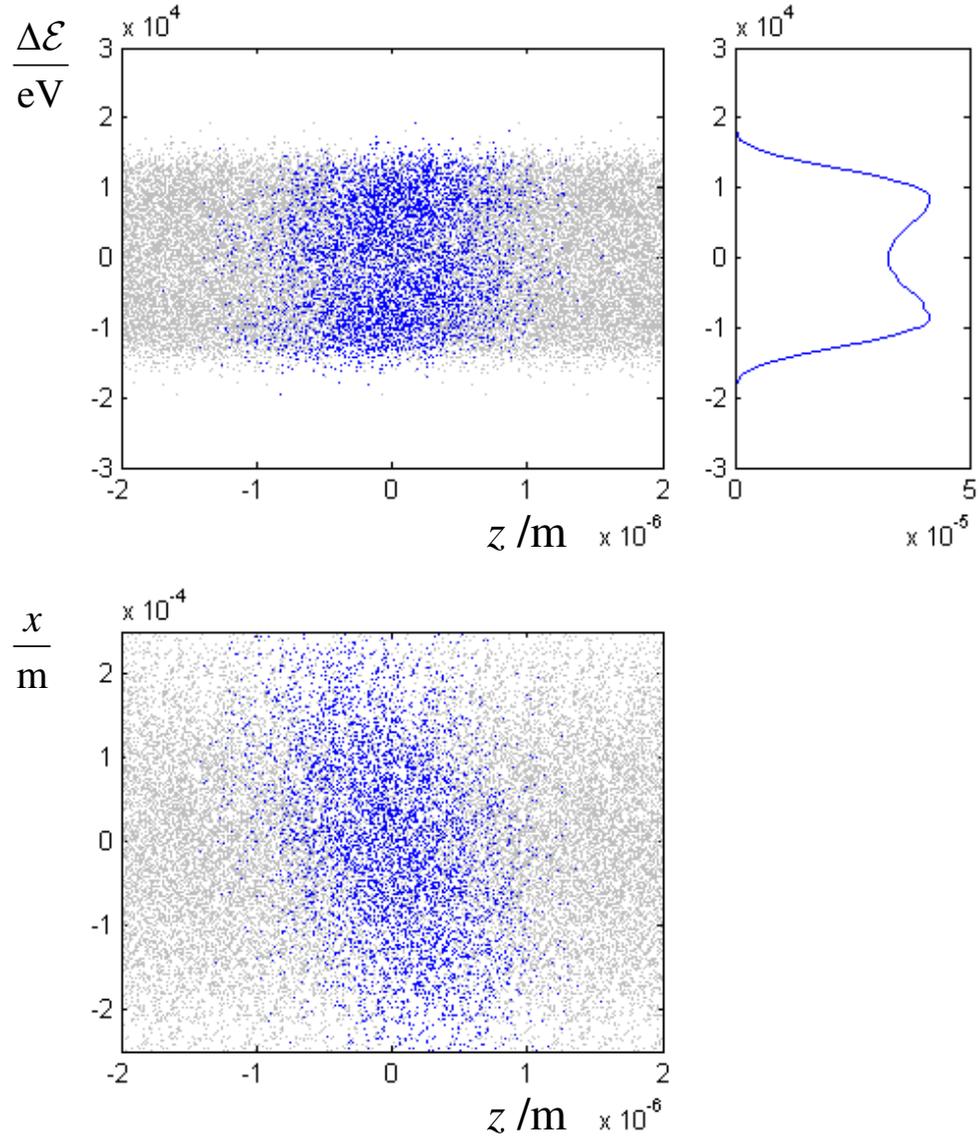} 
\caption{Application C: longitudinal phase space $(\Delta {\cal E},z)$, top view $(x,z)$ and energy spectrum for the position 11 m, short after the chicane.}
\label{LCLS_LH_8keV_11p0}
\end{figure}

At LCLS a laser heater is used to increase the uncorrelated energy spread of the electron beam to counteract a microbunch instability
\cite{LCLS_UB_instab}
. Therefore the electron beam is energy modulated in an undulator that is in the middle of a short chicane, see Fig.~\ref{LCLS_LH}. The periodic modulation directly after the undulator is (in principle) visible in longitudinal phase space. This is different after the dogleg comprising of the last two chicane magnets: the angular divergence before the dogleg, due to emittance, causes spatial longitudinal divergence due to the dispersion parameter $r_{52}.$  This $z$-divergence is large compared to modulation wavelength so that the periodic behavior is obscured, see $(\Delta {\cal E},z)$-diagram in Fig.~\ref{LCLS_LH_8keV_11p0}. Similarly, in spatial space appears no pattern of the harmonic excitation, see $(x,z)$-diagram in Fig.~\ref{LCLS_LH_8keV_11p0}. Therefore it was unexpected to measure energy spectra with rms widths that are {\sl not} proportional to amplitudes of the exciting laser field. The position of the spectrometer can be seen in Fig.~\ref{LCLS_LH}. This effect is a clear indication for self effects and has been discussed, explained and estimated in
\cite{Huang LH}
.

\begin{figure}
\includegraphics[width=0.90\textwidth]{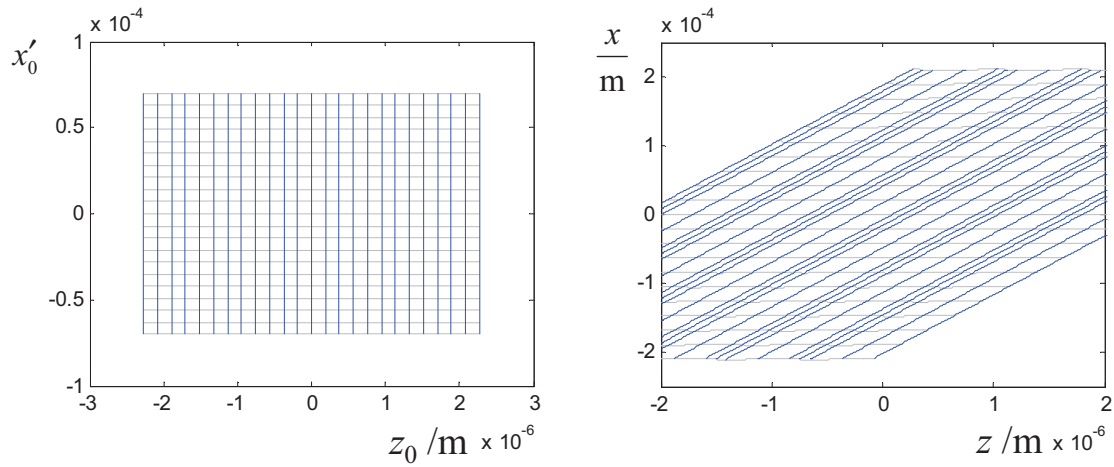} 
\caption{Application C: visualization of transport transformation of $(z_0,x'_0)$, after laser heater undulator to $(x,z)$,at 15.5 m position for particles without offset $x_0$. }
\label{transformation}
\end{figure}

\begin{figure}
\includegraphics[width=0.66\textwidth]{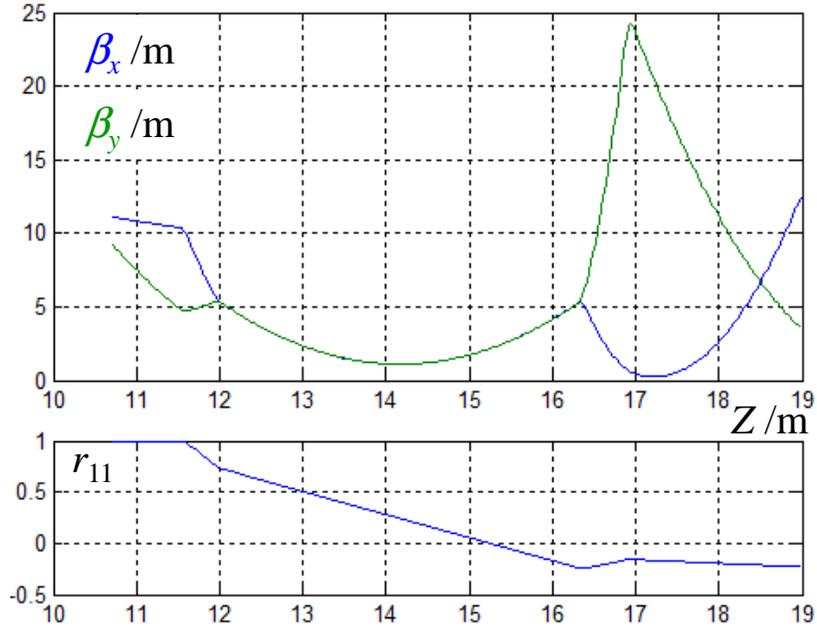} 
\caption{Application C: beta functions and $r_{11}$ from the end of the laser heater undulator to an arbitrary lattice. The plot range is from the exit of the undulator to the beginning of the spectrometer magnet.}
\label{LCLS_LH_optics_and_r11}
\end{figure}

We give a simplified explanation for this effect. The transport equation from the exit of the laser heater undulator to an arbitrary position after the dogleg is
\begin{equation}
\left[\begin{array}{c} x \\ z \end{array}\right] =
\left[\begin{array}{cc} r_{11} & 0 \\ 0 & 1  \end{array}\right]
\left[\begin{array}{c} x_0 \\ z_0 \end{array}\right]
+
\left[\begin{array}{c} r_{12} \\ r_{52} \end{array}\right] x'_0
+
\left[\begin{array}{c} r_{16} \\ r_{56} \end{array}\right] \hat{\delta} \cos (k z_0)
\label{transport_equation}
\: \: ,
\end{equation}
with $x_0, x'_0, z_0, \delta_0=\hat{\delta} \cos (k z_0)$ the phase space coordinates after the undulator and $x, z$ the spatial coordinates (at the arbitrary position) after the dogleg and the relative modulation amplitude $\hat{\delta}$. The energy spread before the undulator is neglected and for the beginning we neglect also the contribution of $x_0$. Fig.~\ref{transformation} illustrates the transformation by Eq.~\ref{transport_equation}: an equidistant orthogonal grid in the $(x'_0,z_0)$-plane is transformed to the $(x,z)$ plane. The chosen numbers are typical for the given setup. It is obvious that the original area has been sheared and modulated and spatial density oscillations have been created.  Also particles with the same initial coordinate $z_0$ but different slope $x'_0$ are distributed horizontally, they are still located on lines. To achieve a homogeneous spatial mixing, a horizontal spread of $x_0$ is necessary that is coupled by $r_{11}$  into the horizontal direction again. Supposed the fluctuation of  $r_{11} x_0$ is sufficiently large (compared to the horizontal distance between lines with $\Delta z_0 = \lambda$), then the density is uniform, as for instance in Fig.~\ref{LCLS_LH_8keV_11p0}. This is different in the vicinity of zero crossings of $r_{11}$ as in Fig.~\ref{LCLS_LH_optics_and_r11} with $r_{11}$ and optical functions $\beta_x, \beta_y$. At these positions appear spatial density fluctuations and space charge fields cause an additional energy modulation, which is correlated with the original modulation.

\begin{figure}
\includegraphics[width=0.80\textwidth]{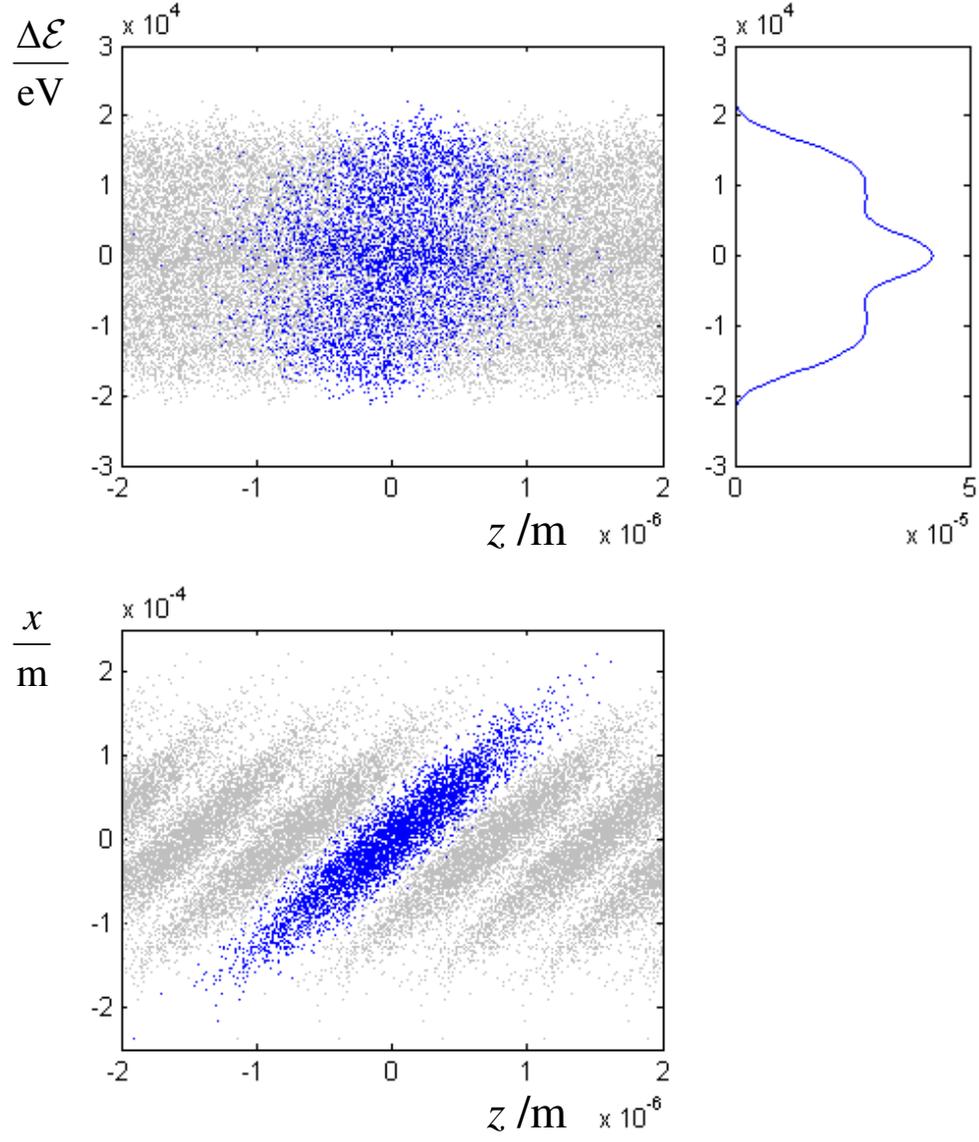} 
\caption{Application C: longitudinal phase space $(\Delta {\cal E},z)$, top view $(x,z)$ and energy spectrum for the position 15.5 m, near the zero crossing of $r_{11}$.}
\label{LCLS_LH_8keV_15p5}
\end{figure}

\begin{figure}
\includegraphics[width=0.80\textwidth]{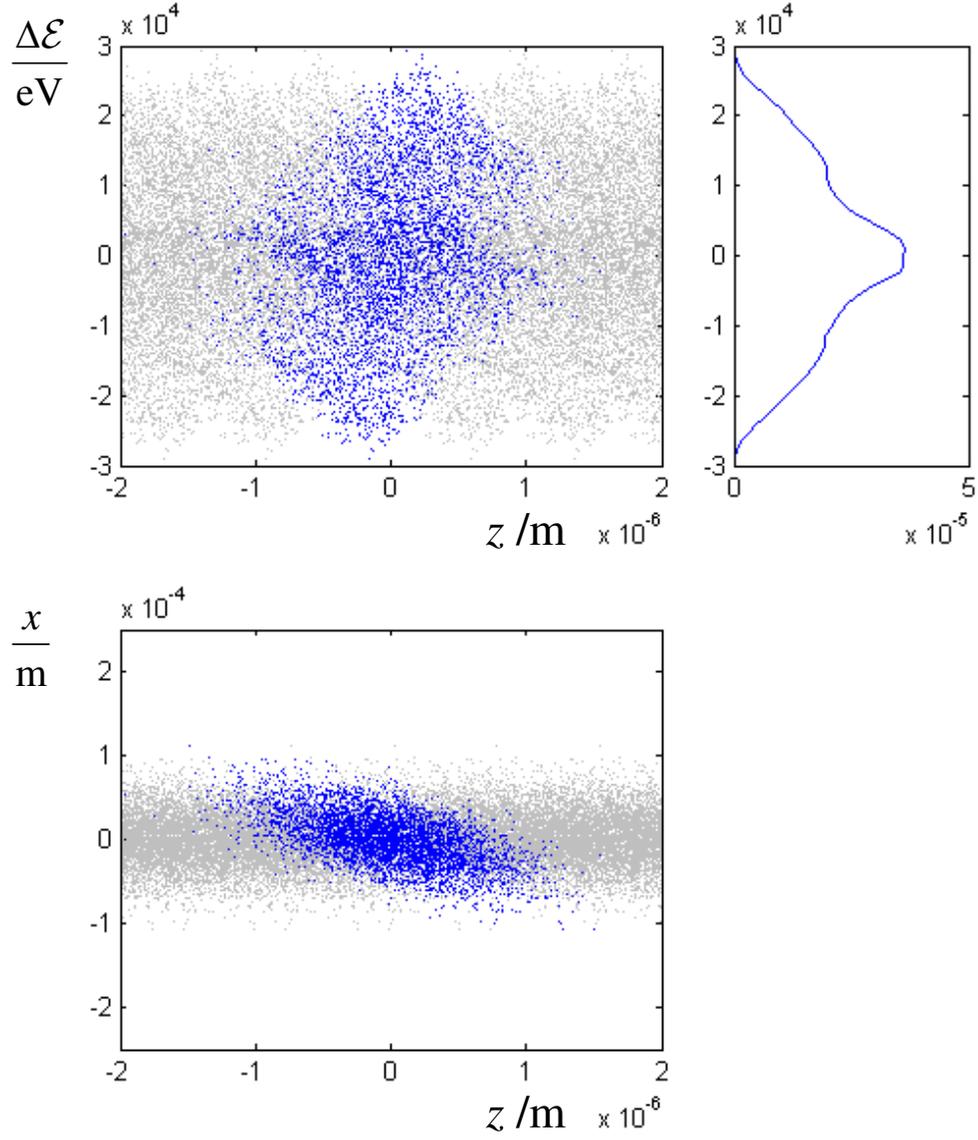} 
\caption{Application C: longitudinal phase space $(\Delta {\cal E},z)$, top view $(x,z)$ and energy spectrum for the position 17.5 m.}
\label{LCLS_LH_8keV_17p5}
\end{figure}

\begin{figure}
\includegraphics[width=0.66\textwidth]{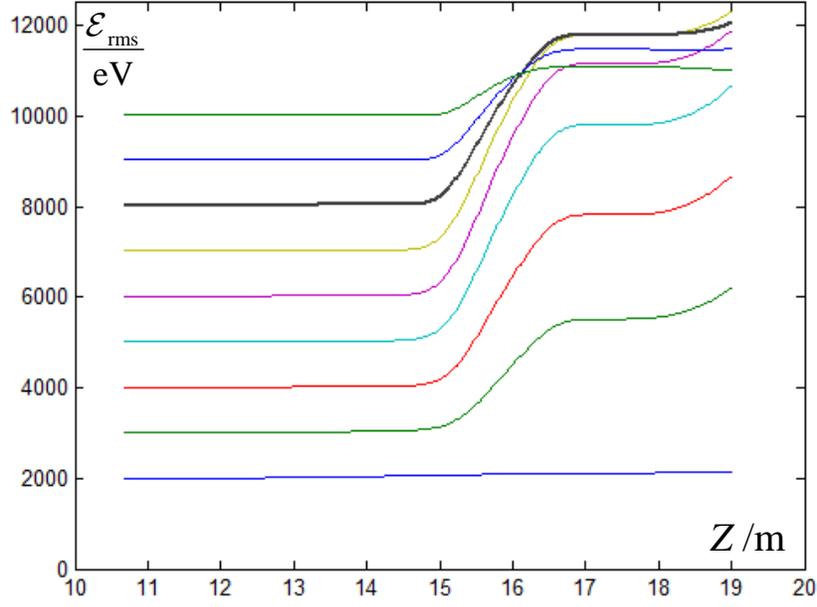} 
\caption{Application C: energy spread after the laser heater chicane as function of the beamline coordinate. The initial spread varies due to different laser modulation.}
\label{LCLS_LH_rms_vs_len}
\end{figure}

\begin{figure}
\includegraphics[width=0.6\textwidth]{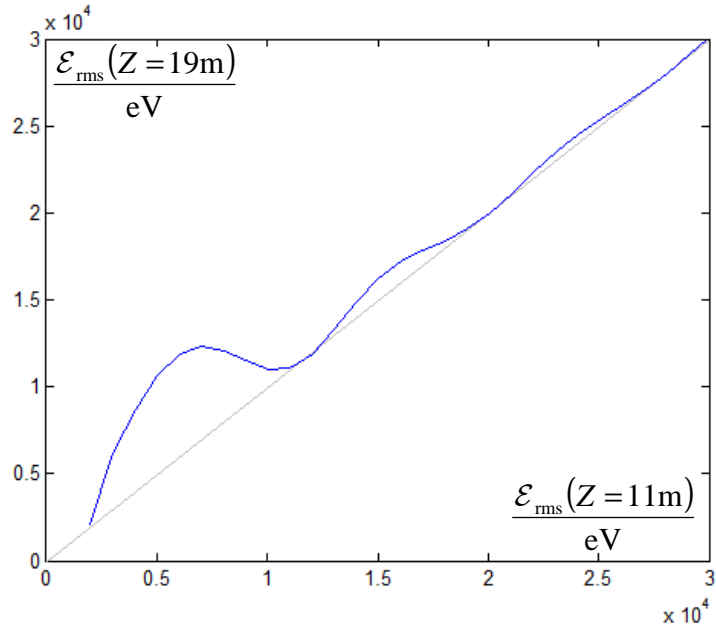} 
\caption{Application C: energy spread at the spectrometer (position 19 m) as function of the spread directly after the laser heater chicane (position 11 m). The initial spread varies due to different laser modulation.}
\label{LCLS_LH_rms_out_vs_rms_in}
\end{figure}

Figs.~\ref{LCLS_LH_8keV_11p0}, \ref{LCLS_LH_8keV_15p5} and \ref{LCLS_LH_8keV_17p5} are results of a numerical simulation with the periodic space charge solver of the setup described in
\cite{Huang LH}
(${\cal E}$ = 135 MeV, $\sigma_{\cal E}$ = 2 keV, peak current 37 A, charge 250 pC, normalized emittance 0.4 $\mu$m, modulation wavelength $\lambda$ = 758 nm).  The beam is heated to $\sigma_{\cal E}$ = 8 keV. The blue particles in the $(\Delta {\cal E},z)$- and $(x,z)$-diagrams are particles that have been in the same longitudinal period during modulation, the gray particles are a periodic repetition.  Fig.~\ref{LCLS_LH_8keV_11p0} refers to a position short after the laser heater chicane. The particles behave as expected by a theory neglecting self effects: the oscillating behavior is not visible, the particles are longitudinally smeared over a length larger than the period $\lambda$. The spectral density agrees with the predicted double bump structure
\cite{LH_double_bump}.
Fig.~\ref{LCLS_LH_8keV_15p5} refers to the position 15.5 m, close to the zero crossing of $r_{11}$. In the spatial diagram the periodic pattern appears as microbunching, although the relative longitudinal position of all the particles is not changed. The space charge fields due to the micro structure changes the particles energy and alters the energy distribution: the spectrum is different and its rms width is increased. At the position 17.5 m shown in Fig.~\ref{LCLS_LH_8keV_17p5}, no microbunching is obvious but the energy distribution is significantly widened to $\sigma_{\cal E} \approx$ 12 keV. Fig.~\ref{LCLS_LH_rms_vs_len} shows the rms energy spread along the beamline for different heating strengths. The blue 2 keV curve is calculated without heating, the thick black line refers to a heating (from initially 2 keV) to 8 keV as in the previous figures. The parasitic heating effect can be clearly seen. The characteristic of this heating described in Fig.~\ref{LCLS_LH_rms_out_vs_rms_in} showing the rms energy spread at the spectrometer entrance versus the initial energy spread. It is remarkable that parasitic heating changes not only the rms width but also the shape of the spectrum, see sub-panels with spectra in Figs.~\ref{LCLS_LH_8keV_11p0}, \ref{LCLS_LH_8keV_15p5} and \ref{LCLS_LH_8keV_17p5}.

This calculation has been performed with 1E6 macro particles per period although the number of electrons per period is 0.55E6. The longitudinal resolution is $\Delta z=\lambda /50$, the transverse resolution is $\Delta x=\Delta_y=\gamma \Delta z$, so that the Lorentz transformed mesh has the same resolution in all directions. The phase space pictures have been computed separately, with a resolution of 20 meshlines per period.

\begin{figure}
\includegraphics[width=0.95\textwidth]{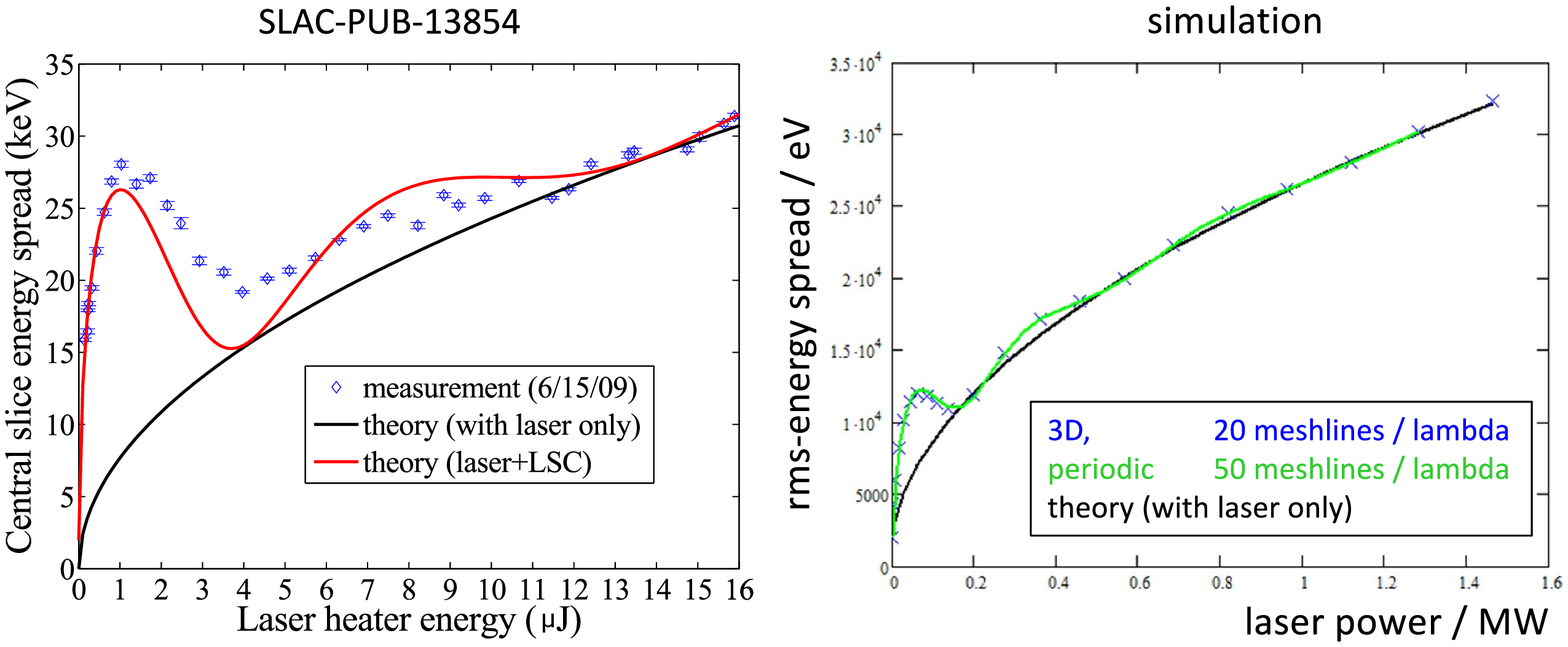} 
\\ \hspace{0.3cm} (a) \hspace{6.7cm} (b)
\caption{(a) Measurement \cite{Huang LH}, (b) simulation with QField-periodic and QField-non-periodic.}
\label{LCLS_LH_meas_vs_calc}
\end{figure}

Fig.~\ref{LCLS_LH_meas_vs_calc} shows the published measurement
\cite{Huang LH}
together with a periodic and non-periodic simulation with QField. Although the shapes agree qualitatively, the
results are not in good agreement: (a) the measured heating is stronger (up to 28 keV where 10 keV are expected without additional heating), (b) the energy values without increased heating are different (here: $\approx$ 11 keV, 20 keV, measured: $\approx$ 16 keV, 28 keV). 


\section{Conclusion}
In this paper we developed a method to calculate the potential and electro-magnetic field of a periodic charge distribution that moves in free space with constant velocity. The charge distribution is three dimensional and periodic in one direction of space. The numerical technique is a modification of the well known particle-mesh-method with fast convolution of the discretized source with the Green's function of one cell. An efficient method has been described to compute the periodic Green's function so that only particles, representing one period, have to be considered for the convolution.

The method has been demonstrated with examples of pure spatial periodicity into the direction of motion. For example (A) a comparison with non-periodic simulations is shown. Although the quasi-periodic part of the bunch is quite short, a good agreement is achieved. A non-periodic simulation for example (B) was beyond the available computational capacities.
For example (C) some points of the curve in Fig.~\ref{LCLS_LH_meas_vs_calc}b have been verified with large effort by non-periodic simulations with lower spatial resolution. The CPU time requirement for each point was about 1000 times larger despite the lower resolution.

Not all capabilities of the method have been demonstrated by these examples. For instance a density modulated beam with energy chirp represents a periodicity in geometric- and momentum-space. Due to longitudinal and transverse dispersion the periodicity vector $\bf{r}_{\rm p}$ changes its length and direction and deviates from the direction of motion. This happens in bunch compressor chicanes.

The periodic approach is numerically much more efficient than non-periodic simulations of bunches with quasi-periodicity, as the number of particles and the volume of the problem domain is reduced by orders of magnitude.


\section*{Acknowledgements}
We thank Winfried decking, Christoph Lechner, Kirsten Hacker, Evgeny Schneidmiller, Mikhail Yurkov, Igor Zagorodnov and Torsten Limberg for 
their interest and stimulating discussions.


\clearpage

%
%
%
%

%
%

\clearpage

%
%
%
%
%
%
%


\begin{thebibliography}{99} \vspace{-2mm}

\bibitem{Hockney} R.~Hockney and J.~Eastwood. Computer Simulation Using Particles. Institute of Physics Publishing, Bristol, 1992.

\bibitem{Qiang} J.~Qiang, S.~Lidia, R.~Ryne and C.~Limborg-Deprey, Three-dimensional quasistatic model for high brightness beam dynamics simulation. Phys. Rev. ST Accel. Beams, 9(4):044202, Apr 2006.

\bibitem{Lawson}J.~Lawson: The Physics of Charged-Particle Beams. Clarendon Press, Oxford 1988.
\bibitem{GSSY_plasma}G.~Geloni, E.~Saldin, E.~Schneidmiller, M.~Yurkov: Benchmark of Astra with Analytical Solution for the Longitudinal Plasma Oscillation Problem. Proceedings FEL Conference 2004.
\bibitem{SPIE_Prag} M.~Dohlus, E.~Schneidmiller, M.~Yurkov, C.~Henning; F.~Gruener: Longitudinal space charge amplifier driven by a laser-plasma accelerator. Proc. SPIE 8779 87791T, 2013.

\bibitem{FLASH} S.~Schreiber: First Lasing in the Water Window with 4.1 nm at FLASH. Proceedings of FEL 2011, Shanghai, China.
\bibitem{Hacker} K.~Hacker et al.: Measurements and Simulations of Seeded Electron Microbunches with Collective Effects. To be published.
\bibitem{SASE_Sup} C.~Lechner, et. al.: Experimental verification of controlling FEL gain by laser-induced microbunching instability, in preparation.

\bibitem{LCLS} P. Emma et al.: First Lasing of the LCLS x-ray FEL at 1.5 A. Proceedings of the 2009 IEEE Particle Accelerator Conference, Vancouver, BC, Canada, May, 2009.
\bibitem{Huang LH} Z.~Huang: Measurements of the LCLS laser heater and its impact on the
x-ray FEL performance. Phys. Rev. ST Accel. Beams 13, 020703, February 2010.

\bibitem{Floettmann} K.~Fl\"ottmann: ``ASTRA'', DESY, Hamburg, \path|www.desy.de/~mpyflo|, 2000.
\bibitem{Saldin-1d-impedance} G.~Geloni, E.~Saldin, E.~Schneidmiller and M.~Yurkov: Longitudinal wake field for an electron beam accelerated through an ultrahigh field gradient. Nucl. Instrum. and Methods, A578, 2007.

\bibitem{LCLS_UB_instab} Z.~Huang, J.~Wu, T.~Shaftan: Microbunching Instability due
to Bunch Compression. SLAC-PUB-11597, December 2005.

\bibitem{LH_double_bump} Z.~Huang, M.~Borland, P.~Emma, J.~Wu, C.~Limborg, G.~Stupakov, J.~Welch: Suppression of microbunching instability in the linac coherent light source. Phys. Rev. ST Accel. Beams 7, 074401, 2004.


\end{thebibliography}
\end{document}